\documentclass[12pt,a4paper]{article}

\usepackage[utf8]{inputenc}
\usepackage[T1]{fontenc}
\usepackage{amsmath,amssymb}
\usepackage{graphicx}
\usepackage[version=4]{mhchem}
\usepackage{hyperref}
\usepackage{natbib}
\usepackage{geometry}
\usepackage{float}
\usepackage{subcaption}
\usepackage{booktabs}
\usepackage{array}
\usepackage{makecell}

\title{Composition-Dependent Plasmon-Enhanced Emission in Lead-Free \ce{Cs3Cu2X5} Halides: A DFT--FDTD Study}

\author{
Shoumik Debnath$^{a}$, Sudipta Saha$^{a}$, Khondokar Zahin$^{a}$, Ying Yin Tsui$^{b}$, \\
and Md. Zahurul Islam$^{a,*}$ \\[2ex]
\small $^{a}$Department of Electrical and Electronic Engineering, \\
\small Bangladesh University of Engineering and Technology, Dhaka 1000, Bangladesh \\[0.5ex]
\small $^{b}$Department of Electrical and Computer Engineering, \\
\small University of Alberta, Edmonton, AB T6G 2H5, Canada \\[0.5ex]
\small $^{*}$Corresponding author. E-mail: mdzahurulislam@eee.buet.ac.bd
}

\date{}

\begin{document}

\maketitle

\begin{abstract}
Lead-free \ce{Cs3Cu2X5} (X = Cl, Br, I) halides show high photoluminescence quantum yields (PLQY) and good ambient stability, yet LEDs based on these materials still suffer from poor optical outcoupling. In this work, we combine density functional theory (DFT) and finite-difference time-domain (FDTD) simulations to optimize plasmonic enhancement in \ce{Cs3Cu2X5} LEDs using composition-specific optical constants. First-principles calculations provide wavelength-dependent refractive index and extinction coefficient data for each halide. These values are then used in FDTD to model a complete device stack with Ag/\ce{SiO2} core-shell nanostructures. Out of the three halides, \ce{Cs3Cu2Cl5} performs best with 4.4$\times$ Purcell enhancement and 30\% light extraction using optimized nanorods. The chloride outperforms the others due to its lower refractive index ($n \approx 1.9$). \ce{Cs3Cu2Br5} has the highest spectral overlap (95.5\%) but only moderate extraction efficiency (26\%) because of increased optical confinement. \ce{Cs3Cu2I5} requires a nanosphere geometry and shows limited performance (10\% extraction) despite reasonable Purcell enhancement. Distance-dependent analysis shows that optimal emitter-plasmon separation varies with composition, ranging from 8--12~nm for \ce{Cs3Cu2Br5} to approximately 15~nm for \ce{Cs3Cu2Cl5}. These results provide composition-specific design guidelines for plasmon-enhanced lead-free LEDs.
\end{abstract}

\section{Introduction}

Metal halide perovskite LEDs have advanced rapidly over the past decade.Red emitters now hit 28\% external quantum efficiency (EQE)~\cite{Kong2024}, green devices reach 30\%~\cite{Li2024a}, and blue recently crossed 23\%~\cite{Nong2024}. These numbers rival state-of-the-art OLEDs.\cite{Cho2024,Li2024b} The performances are achieved due to high absorption coefficients, tunable emission across visible to near-infrared wavelengths, narrow linewidths below 20 nm, and high photoluminescence quantum yields.\cite{Ma2025,Xiao2024,Yang2022} However, two fundamental problems remain. First, lead toxicity blocks commercialization.\cite{Lopez2024,Zhang2021a,Chen2021} Second, even high-yield emitters lose roughly 80\% of generated photons to waveguide and substrate modes.\cite{Zhao2023,Liu2024a,Rahimi2024} Simply replacing lead will not solve the photon extraction bottleneck.

Early lead-free efforts targeted tin and germanium perovskites because \ce{Sn^{2+}} and \ce{Ge^{2+}} mimic the \ce{Pb^{2+}} valence configuration.\cite{Lopez2024,Kar2021} Tin-based PeLEDs have reached EQEs near 20\%.\cite{Xiao2024} But both cations oxidize rapidly to tetravalent states under ambient conditions, creating deep-level defects that kill emission within hours.\cite{Xiao2024,Lopez2024,Wang2024a} Tin is especially unstable due to its low \ce{Sn^{2+}}/\ce{Sn^{4+}} redox potential of +0.15 V.\cite{Wang2024a} For commercial viability, devices need to operate beyond 10,000 hours.\cite{Zhang2025} Neither lead-based nor current lead-free systems meet this requirement alongside cost and safety constraints.

The optical extraction problem persists regardless of the emitter material. Halide perovskites have refractive indices around 2.5,\cite{Cho2024,Rahimi2024,rahman2026hierarchically} which limits outcoupling efficiency to roughly 8\%.\cite{Cho2024} Total internal reflection alone accounts for about 56\% of radiative power loss.\cite{Liu2024a,Rahimi2024} Fixing these losses requires accurate wavelength-dependent optical constants, specifically the refractive index ($n$) and extinction coefficient ($k$), which vary with composition and emission energy.\cite{Shi2018,Zhang2020} Most optical models use generic $n$ and $k$ datasets that cause spectral mismatch between photoluminescence and cavity resonances.\cite{Shi2018,Futscher2017} Without composition-specific dispersion data, device optimization remains empirical rather than predictive.

These challenges share a common root. Material development and photonic engineering have been treated as separate problems.\cite{Chen2024} Synthetic chemists optimize composition for stability without comprehensive optical characterization.\cite{Zainab2023} Device engineers use placeholder optical constants that may not match the actual material.\cite{Akhter2015} The result is incomplete solutions. Materials that resist degradation can still suffer from poor photon extraction. High-PLQY emitters placed in optically mismatched environments underperform. Real progress requires tackling both issues together.

Copper-based halides with the \ce{Cs3Cu2X5} stoichiometry (X = Cl, Br, I) offer a promising path forward. Unlike conventional three-dimensional ABX$_3$ perovskites, \ce{Cs3Cu2X5} has a zero-dimensional crystal structure consisting of isolated \ce{[Cu2X5]^{3-}} bi-octahedral dimers separated by cesium cations.\cite{Yang2023,Li2025a} This arrangement confines excitons to individual molecular units. The monovalent \ce{Cu+} (3d$^{10}$) configuration provides a closed-shell electronic structure that resists oxidation, eliminating the instability pathways that plague \ce{Sn^{2+}} and \ce{Ge^{2+}} systems.\cite{Lian2020,Li2020,Yue2019} As a result, \ce{Cs3Cu2X5} films maintain structural and optical integrity for weeks under ambient conditions without encapsulation.\cite{Li2025a,Gao2025} Recent work has reported PLQYs approaching 95\%,\cite{Lian2020,Jiang2023} with emission tunable from blue to near-infrared through halide substitution.\cite{Li2020,Wang2025} \ce{Cs3Cu2Cl5} emits blue-green around 440 to 520 nm,\cite{Jiang2023,Liu2021} \ce{Cs3Cu2Br5} emits green-yellow around 540 to 580 nm,\cite{Wang2025} and \ce{Cs3Cu2I5} emits in the green region around 530 to 550~nm.\cite{Li2020,Wen2024} Strong electron-phonon coupling within the isolated dimers produces self-trapped excitons with broadband emission and large Stokes shifts around 200 to 600 meV.\cite{Lian2020,Zhang2021b} Still, PeLEDs using these materials show modest EQEs around 1 to 5\%.\cite{Kim2023} Chemical stability alone does not guarantee good device performance.

Addressing the optical extraction challenge requires computational methods that bridge material-scale dispersion and device-scale electromagnetics.\cite{Akhter2015} First-principles DFT calculations using hybrid functionals like HSE06 with spin-orbit coupling provide accurate band structures and dielectric functions.\cite{Li2024c,Xiong2025} The resulting wavelength-dependent $n$ and $k$ spectra capture intrinsic material dispersion without relying on extrapolated values.\cite{Ramirez2019} However, DFT describes bulk properties and cannot model electromagnetic field dynamics within multilayer device architectures.\cite{Yang2023,Chen2024} The FDTD method fills this gap by solving Maxwell's equations numerically in space and time.\cite{Rahimi2024,Tabibifar2024} Importing DFT-derived optical constants into FDTD device models allows physically consistent evaluation of Purcell factor, light extraction efficiency, spectral overlap, and far-field profiles.\cite{Ma2021}

Beyond passive optical engineering, plasmonic coupling through embedded metallic nanostructures offers active enhancement of light extraction.\cite{Zhao2023} Noble-metal nanoparticles support localized surface plasmon resonances that concentrate electromagnetic fields at the nanoscale and accelerate radiative recombination.\cite{Bueno2025} Gold nanorods combine excellent chemical stability with tunable plasmon resonances. Their longitudinal mode can be engineered through aspect-ratio control to align with \ce{Cs3Cu2X5} emission, achieving strong spectral overlap and Purcell enhancement.\cite{Ain2024,Li2025b} But accurate design of plasmon-emitter coupling requires composition-specific optical constants to predict resonance alignment at the nanoscale.

\begin{figure}[t]
\centering
\includegraphics[width=0.7\textwidth]{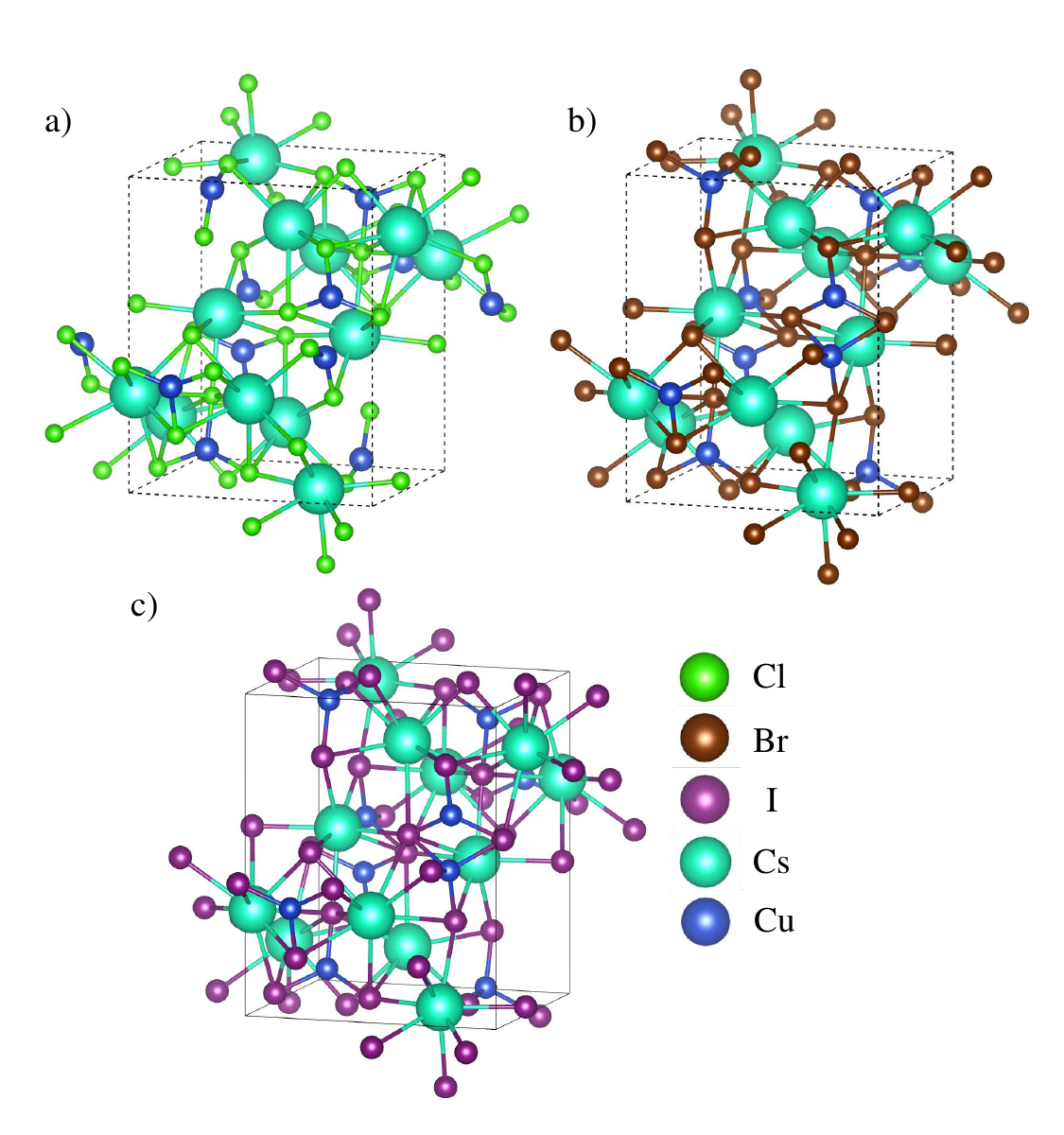}
\caption{Crystal structure of \ce{Cs3Cu2X5} (a) X = Cl, 
(b) X = Br, and (c) X = I. 
The green, brown, violet, sky blue, and blue atom colors correspond to Cl, Br, I, Cs, and Cu atoms, respectively.}
\label{fig:crystal}
\end{figure}

Despite growing interest in lead-free perovskites, systematic studies linking halide composition, optical constants, and plasmonic performance remain scarce.\cite{Wen2024b} Prior research has mostly focused on structural stability or bandgap tuning for photovoltaics.\cite{Yang2023} Most plasmonic simulations still use generalized $n$ and $k$ datasets, undermining the accuracy of enhancement predictions.\cite{Googasian2023} The quantitative connection between halide composition, optical dispersion, plasmon resonance alignment, and device-level performance has not been established for \ce{Cs3Cu2X5} systems.\cite{Wen2024b}

Here, we develop a fully integrated DFT-FDTD framework that establishes quantitative links between halide composition, wavelength-dependent optical properties, and plasmonic enhancement in \ce{Cs3Cu2X5}-based PeLEDs. Composition-specific $n$ and $k$ spectra derived from DFT are directly implemented in FDTD simulations that model emission enhancement, nanorod-emitter coupling, and far-field radiation. We further optimize Ag-nanorod geometries to maximize light extraction efficiency (LEE), Purcell factor, and spectral overlap. This integrated approach combines the chemical stability of copper-based halides with engineered photonic control, providing a predictive design pathway from atomic composition to device-level light extraction.

\begin{figure}[h]
 \centering
 \includegraphics[width=0.95\textwidth]{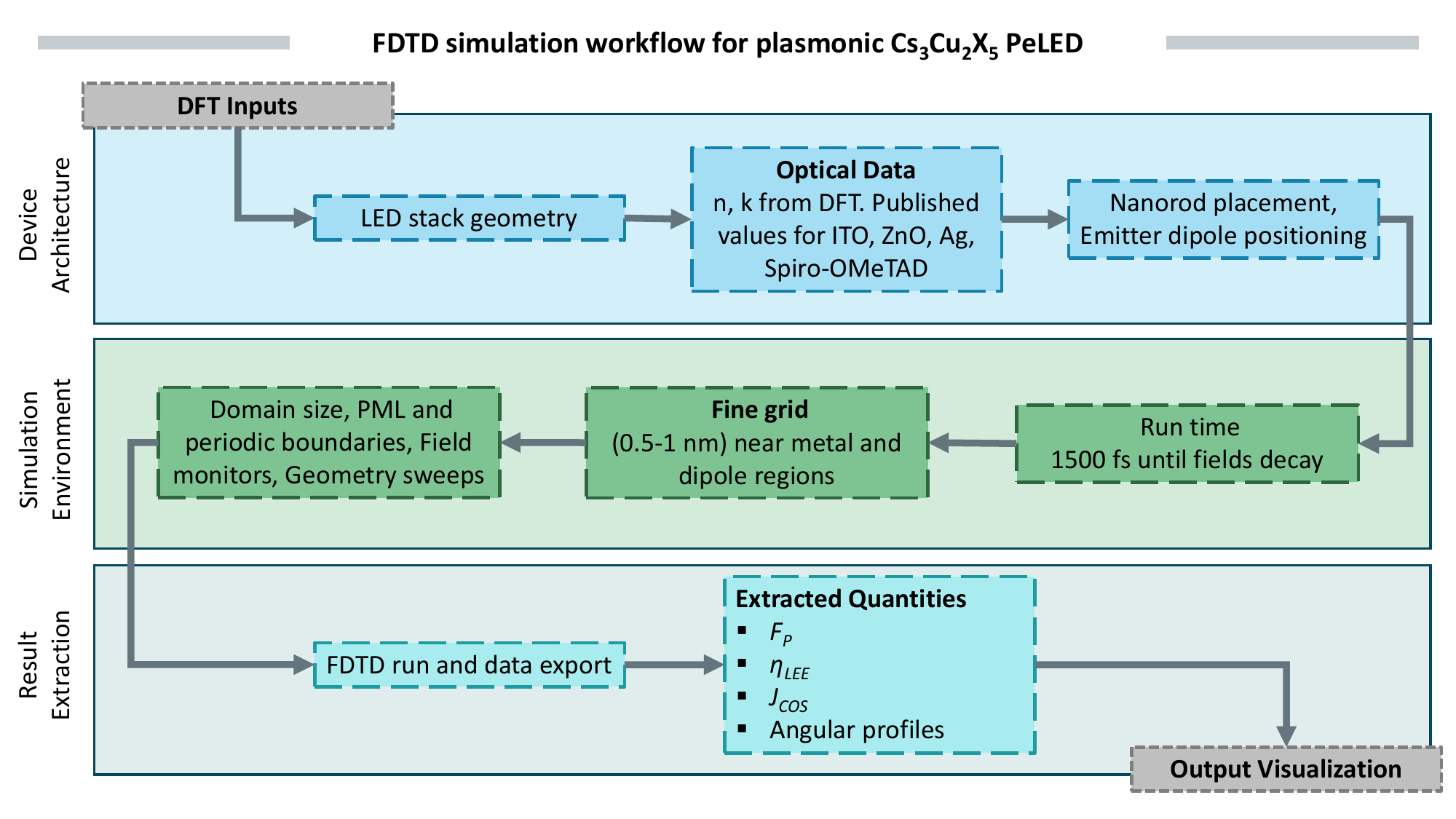}
 \caption{Overview of the DFT-FDTD modeling approach adopted in this work. Optical constants from first-principles calculations feed into device-level electromagnetic simulations that extract key performance metrics.}
 \label{fig:methodology}
\end{figure}

\section{Methodology and Theoretical Background}

\subsection{DFT Methodology}

We carried out all density functional theory calculations using the CASTEP code. We treated exchange-correlation effects with the PBE functional within the generalized gradient approximation framework. Ultrasoft pseudopotentials described the interactions between ions and electrons. We treated the following valence electron configurations explicitly: cesium 6s$^1$, copper 3d$^{9}$4s$^2$, chlorine 3s$^2$3p$^5$, bromine 4s$^2$4p$^5$, and iodine 5s$^2$5p$^5$. We expanded the electronic wavefunctions using a plane-wave basis set with a kinetic energy cutoff of 450~eV. For Brillouin zone integration, we used a $6 \times 6 \times 6$ Monkhorst-Pack k-point grid.

We obtained starting structures from experimentally determined orthorhombic unit cells in the \textit{Pnma} space group. Each unit cell contained 40 atoms. We relaxed both lattice vectors and atomic positions for all three halide compositions using the BFGS optimization algorithm. We continued the relaxation until total energy changes dropped below $1 \times 10^{-5}$~eV/atom, residual forces fell under 0.03~eV/\AA, and atomic displacements stayed within 0.001~\AA. These tight thresholds ensured well-converged ground state geometries.

After obtaining relaxed structures, we computed electronic band structures and density of states. We kept the same k-point mesh and smearing parameters from the ground state runs for consistency. We plotted band dispersions along the standard high-symmetry path for orthorhombic crystals, going through $\Gamma$, Z, Y, T, S, X, U, and R points in sequence. We set the Fermi level to zero energy in all plots. We then calculated optical properties from the frequency-dependent dielectric function. We evaluated the dielectric response for photon energies spanning 0 to 20~eV using the independent particle approximation. This approach captures interband transitions that govern absorption and refractive index behavior in the visible and ultraviolet ranges.

\begin{figure}[htbp]
    \centering
    \begin{minipage}[b]{0.8\textwidth}
        \centering
        \includegraphics[width=\textwidth]{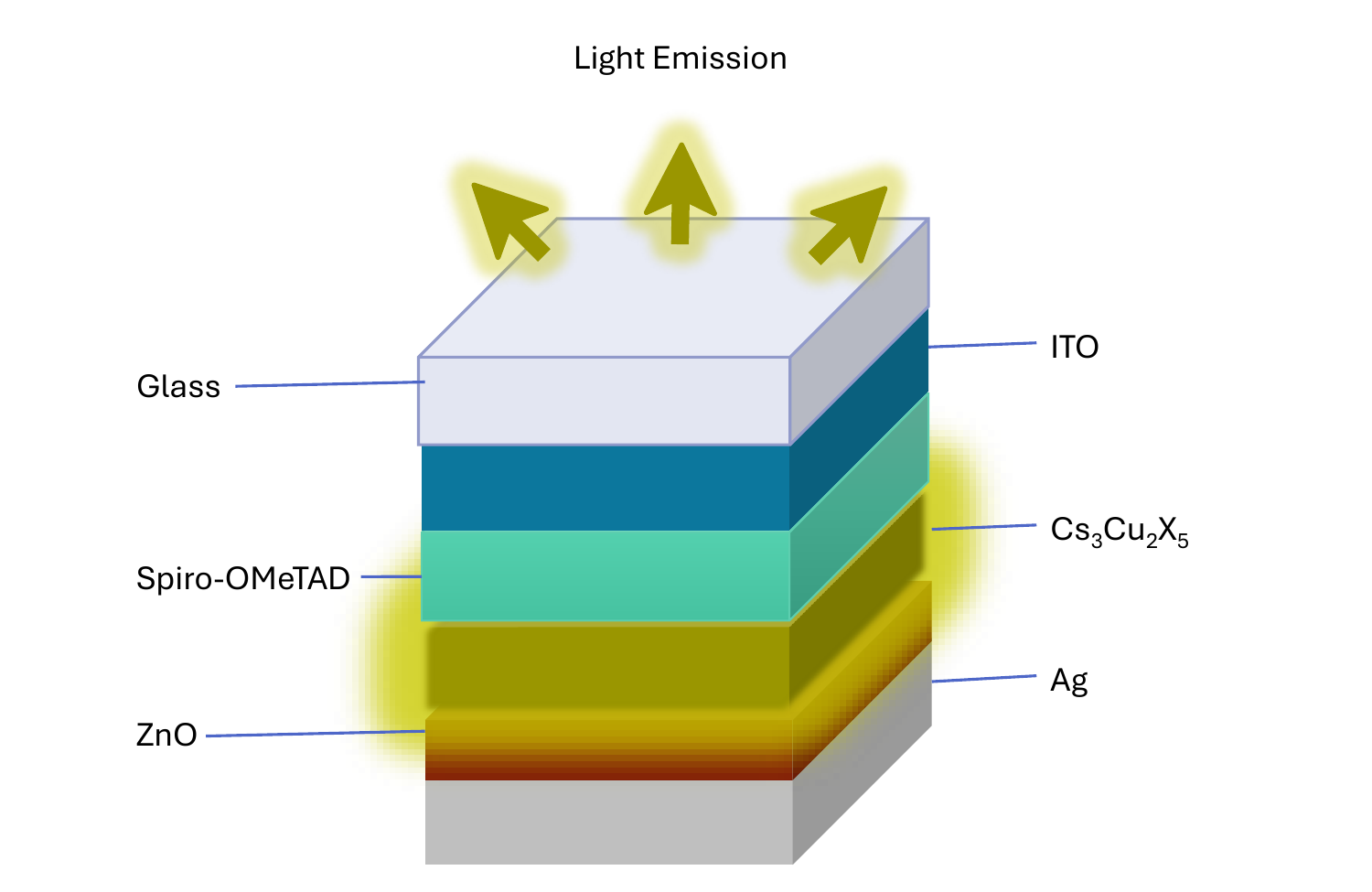}
        (a)
    \end{minipage}
    
    \vspace{0.5cm}
    
    \begin{minipage}[b]{0.8\textwidth}
        \centering
        \includegraphics[width=\textwidth]{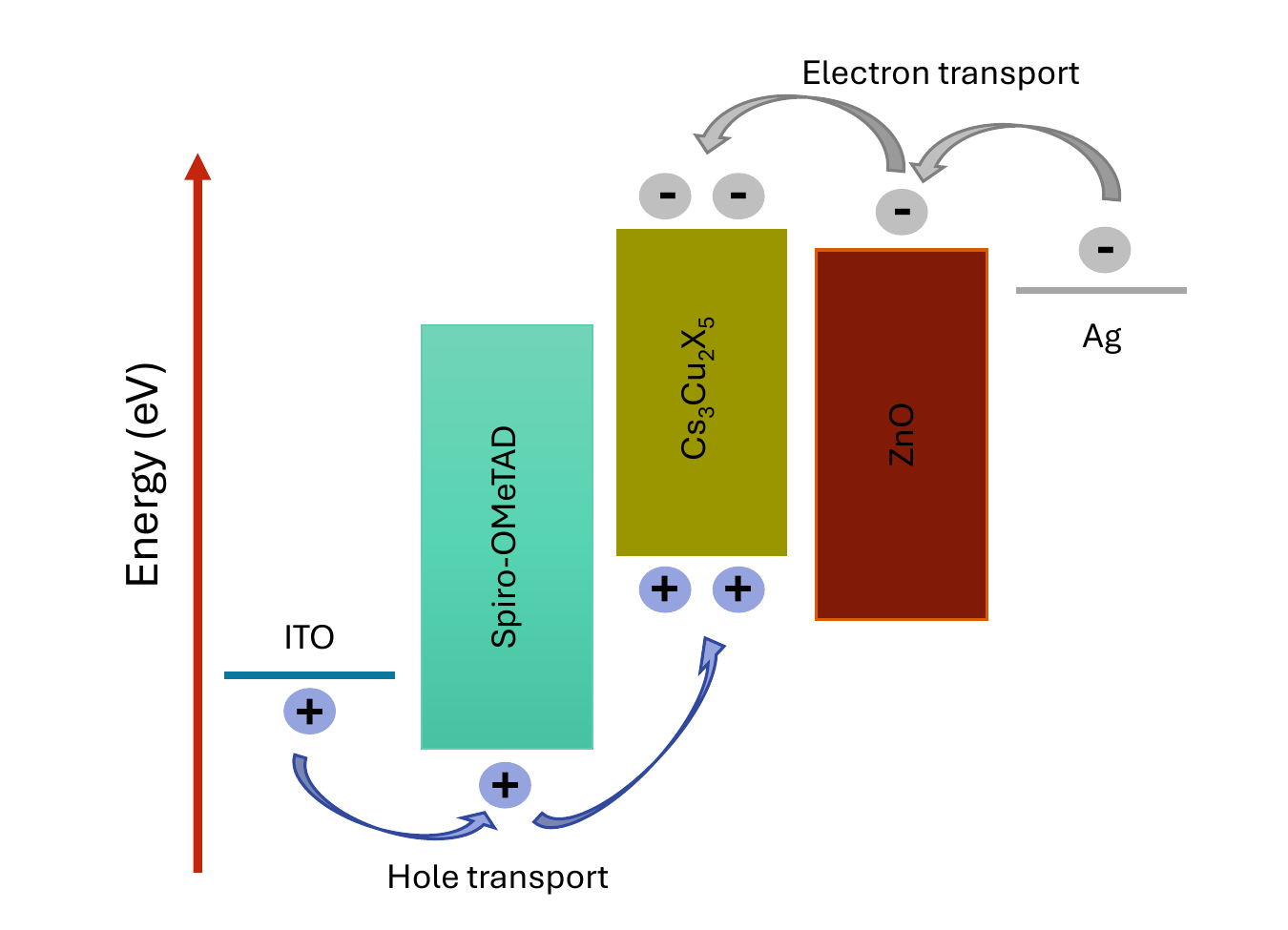}
        (b)
    \end{minipage}
  \caption{(a) Schematic of the simulated PeLED stack. (b) Energy band alignment showing hole injection from ITO through Spiro-OMeTAD and electron injection from Ag through ZnO into the perovskite active layer.}
  \label{fig:stack}
\end{figure}

\subsection{FDTD Simulation Setup}

We modeled the optical behavior of plasmonic \ce{Cs3Cu2X5} LEDs using three-dimensional FDTD simulations in Lumerical (Ansys Inc.).\cite{Taflove2005} Figure~2 shows the computational workflow.

We designed the device stack to match experimentally reported LED architectures.\cite{Zhou2021} Starting from the bottom, we used 100~nm of ITO as the anode. On top of that we placed a 35~nm Spiro-OMeTAD hole transport layer. The 50~nm \ce{Cs3Cu2X5} active layer came next, followed by 40~nm of ZnO as the electron transport layer. A 100~nm Ag film served as the top cathode. We imported the refractive index and extinction coefficient spectra for each \ce{Cs3Cu2X5} composition directly from our DFT results. For the other device layers, we used optical constants from published literature.\cite{Brivio2014,Palik1985,Alnuaimi2016}

We placed an electric dipole source inside the perovskite layer to represent spontaneous emission. The dipole radiated across visible wavelengths matching the photoluminescence range of \ce{Cs3Cu2X5}. We positioned this source a few nanometers away from the plasmonic nanostructure surface to capture near-field coupling effects.\cite{Purcell1946}

We chose boundary conditions to balance physical accuracy with computational efficiency. Along the lateral $x$ and $y$ directions, we applied periodic boundaries to simulate an extended array of emitters. In the vertical $z$ direction, we used perfectly matched layers to absorb outgoing waves and prevent artificial reflections from the simulation edges.\cite{Berenger1994} We used a non-uniform spatial mesh throughout the simulation domain. We refined the grid spacing to 0.5--1~nm near the metal nanostructure and dipole region where field gradients are steepest. We increased the mesh spacing away from these areas to reduce memory use. We ran each simulation for 1500~fs, which gave the electromagnetic fields enough time to decay fully before we performed Fourier transforms to extract spectral data.

\subsection{Plasmonic Nanostructure Design}

We selected the nanoparticle geometry based on spectral overlap optimization between the localized surface plasmon resonance (LSPR) and the emission wavelength of each perovskite composition. Nanorods offer a tunable longitudinal plasmon mode through aspect-ratio control, making them suitable for targeting specific emission wavelengths. Nanospheres, in contrast, have a fixed dipolar resonance governed primarily by particle size and the surrounding dielectric environment. We evaluated both geometries for each \ce{Cs3Cu2X5} variant and selected the structure with superior spectral alignment.

We positioned all nanostructures near the ZnO/perovskite interface (Fig.~4) with a core-shell architecture. We fixed the \ce{SiO2} shell thickness at 5~nm to position the emitter within the strong near-field enhancement regime while avoiding non-radiative energy transfer to the metal, which dominates below approximately 4~nm and leads to net quenching~\cite{anger2006enhc}. The shell also serves as a passivation layer to prevent charge carrier quenching. We used the Palik dataset for the frequency-dependent dielectric function of silver~\cite{Palik1985}.

We optimized nanorod dimensions through parametric sweeps, varying the diameter from 8 to 25~nm and the length from 30 to 70~nm. We swept the nanosphere radius similarly to identify optimal coupling conditions. The optimization target was maximum spectral overlap between the LSPR and the composition-dependent emission peak of each \ce{Cs3Cu2X5} variant.

\begin{figure}[h]
 \centering
 \includegraphics[width=0.8\textwidth]{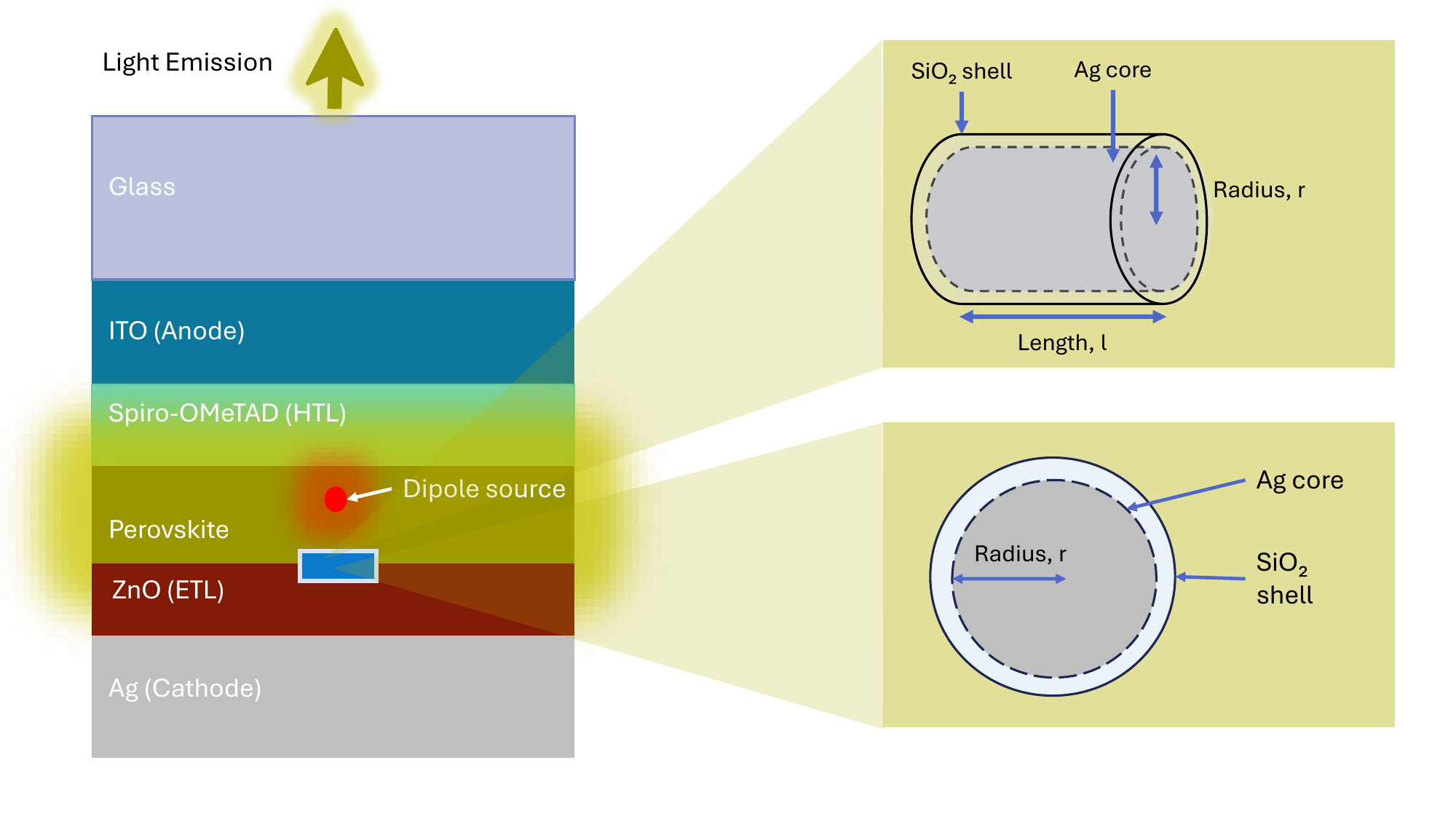}
 \caption{Placement of the dipole source and plasmonic nanostructures within the LED stack. A core--shell Ag/\ce{SiO2} nanorod (variable radius 8--25 nm, length 30--70 nm) is used for \ce{Cs3Cu2Cl5} and \ce{Cs3Cu2Br5}, while a nanosphere geometry is employed for \ce{Cs3Cu2I5}. The 5 nm \ce{SiO2} shell prevents non-radiative quenching while maintaining near-field coupling.}
 \label{fig:nanorod}
\end{figure}

\subsection{Compositional Analysis Strategy}

We ran systematic simulations across the \ce{Cs3Cu2X5} (X = Cl, Br, I) compositional series, both with and without the plasmonic nanostructure. For each halide composition, we used the corresponding DFT-derived $n$ and $k$ spectra in the FDTD model.

This dual-simulation approach isolates the influence of halide-induced optical dispersion changes from plasmonic coupling effects.\cite{Chen2019} We normalized all extracted optical parameters to their non-plasmonic counterparts to determine the enhancement introduced by the nanostructure for each composition.

From the computed field data, we extracted Purcell factor, light extraction efficiency, near-field intensity distributions, and far-field angular patterns.

\subsection{Performance Metrics}

\textbf{Purcell Factor.} The Purcell factor ($F_P$) quantifies the enhancement 
of spontaneous emission rate when an emitter is placed in a modified photonic 
environment compared to free space.\cite{Purcell1946} It arises from changes 
in the local density of optical states (LDOS) experienced by the dipole emitter. 
In FDTD simulations, $F_P$ is calculated as the ratio of total power radiated 
by the dipole in the device structure ($P_{\text{cav}}$) to that in a 
homogeneous reference medium of equivalent refractive index 
($P_0$)\cite{Taflove2005}
\begin{equation}
F_P = \frac{P_{\text{cav}}}{P_0}
\label{eq:purcell_fdtd}
\end{equation}
This approach is valid because the emission rate is proportional to the LDOS, 
and the LDOS is proportional to the power emitted by the source. Values greater 
than unity indicate enhanced radiative recombination. Plasmonic nanostructures 
can produce large Purcell factors by concentrating electromagnetic fields near 
the emitter.\cite{Gu2020}

\textbf{LEE.} The LEE represents the fraction of emitted photons that escape the device into the 
far field rather than being trapped by waveguide modes, substrate absorption, 
or total internal reflection.\cite{Raypah2022} LEE is defined as
\begin{equation}
\eta_{\text{LEE}} = \frac{P_{\text{out}}}{P_{\text{total}}}
\label{eq:lee}
\end{equation}
where $P_{\text{out}}$ is the power radiated into the upper hemisphere and 
$P_{\text{total}}$ is the total power emitted by the dipole source. In FDTD 
simulations, $P_{\text{out}}$ is calculated by integrating the Poynting vector 
over a far-field monitor placed above the device structure. High refractive 
index contrast at layer interfaces typically limits LEE in perovskite devices 
to below 20\% without optical engineering.\cite{Rahimi2024, Zhao2023} Plasmonic 
structures can redirect trapped modes into propagating radiation, thereby 
improving outcoupling. The LEE enhancement factor is defined as the ratio of 
LEE with and without the plasmonic nanostructure.

\textbf{Spectral Overlap.} Efficient plasmon-emitter coupling requires spectral 
alignment between the emitter photoluminescence and the LSPR of the nanostructure.\cite{Wurtz2007} The degree of 
overlap determines the strength of near-field interaction and the magnitude 
of Purcell enhancement. To quantify spectral matching independent of amplitude, 
the cosine similarity metric ($J_{\cos}$) is employed
\begin{equation}
J_{\cos} = \frac{\int S(\lambda) \, C(\lambda) \, d\lambda}{\sqrt{\int S^2(\lambda) \, d\lambda} \sqrt{\int C^2(\lambda) \, d\lambda}}
\label{eq:jcos}
\end{equation}
where $S(\lambda)$ is the normalized emission spectrum and $C(\lambda)$ is the 
normalized plasmonic scattering cross-section spectrum. A value of $J_{\cos} = 1$ 
indicates perfect spectral alignment and $J_{\cos} = 0$ indicates no overlap. 
Nanostructure dimensions were optimized to maximize spectral overlap with the 
composition-dependent emission peaks of each \ce{Cs3Cu2X5} variant.

\textbf{Far-Field Radiation Pattern.} The far-field angular distribution 
describes the directional characteristics of light emitted from the 
device.\cite{Mao2021} It is computed by projecting the near-field data to the 
far-field regime through standard electromagnetic transformations involving 
Fourier transformation of the tangential electric field components recorded 
at the detection surface.\cite{Kim2005} Directional emission concentrated 
within a narrow angular cone improves practical light collection and 
utilization. Plasmonic and cavity effects can reshape the far-field pattern, 
concentrating emission toward the surface normal.\cite{Ooi2024}

\section{Results and Discussion}

\subsection{Electrical and Optical Properties of Cs$_3$Cu$_2$X$_5$}

\begin{figure}[h]
 \centering
 \includegraphics[width=0.95\textwidth]{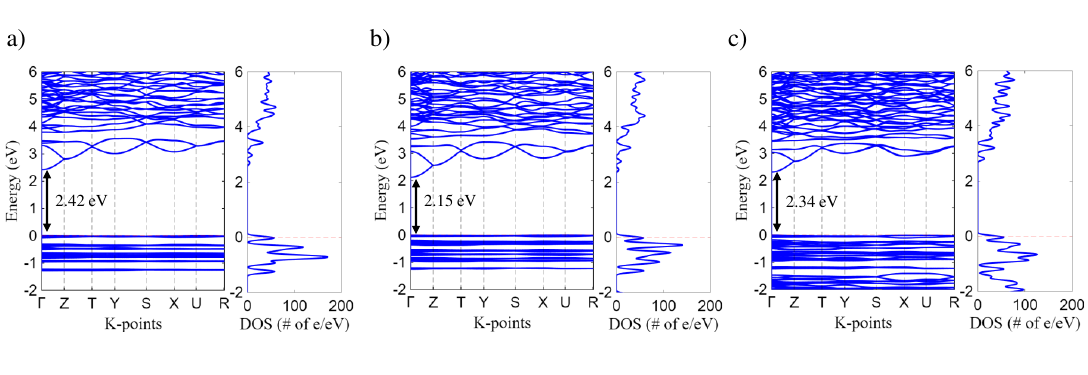}
 \caption{Calculated electronic band structures and total density of states (DOS) of \ce{Cs_3Cu_2X_5} compounds: (a) \ce{Cs_3Cu_2Cl_5}, (b) \ce{Cs_3Cu_2Br_5}, and (c) \ce{Cs_3Cu_2I_5}. The band structures are plotted along the high-symmetry directions of the Brillouin zone, as indicated on the horizontal axis, with the Fermi level set to 0~eV (red dashed line). The corresponding total DOS is shown in the right panel of each subfigure. All compounds exhibit direct band gaps of 2.42~eV, 2.15~eV, and 2.34~eV for \ce{Cs_3Cu_2Cl_5}, \ce{Cs_3Cu_2Br_5}, and \ce{Cs_3Cu_2I_5}, respectively, as marked in the band structure plots.}
 \label{fig:dos}
\end{figure}

\begin{figure}[h]
 \centering
 \includegraphics[width=0.95\textwidth, trim=0 130 0 100, clip]{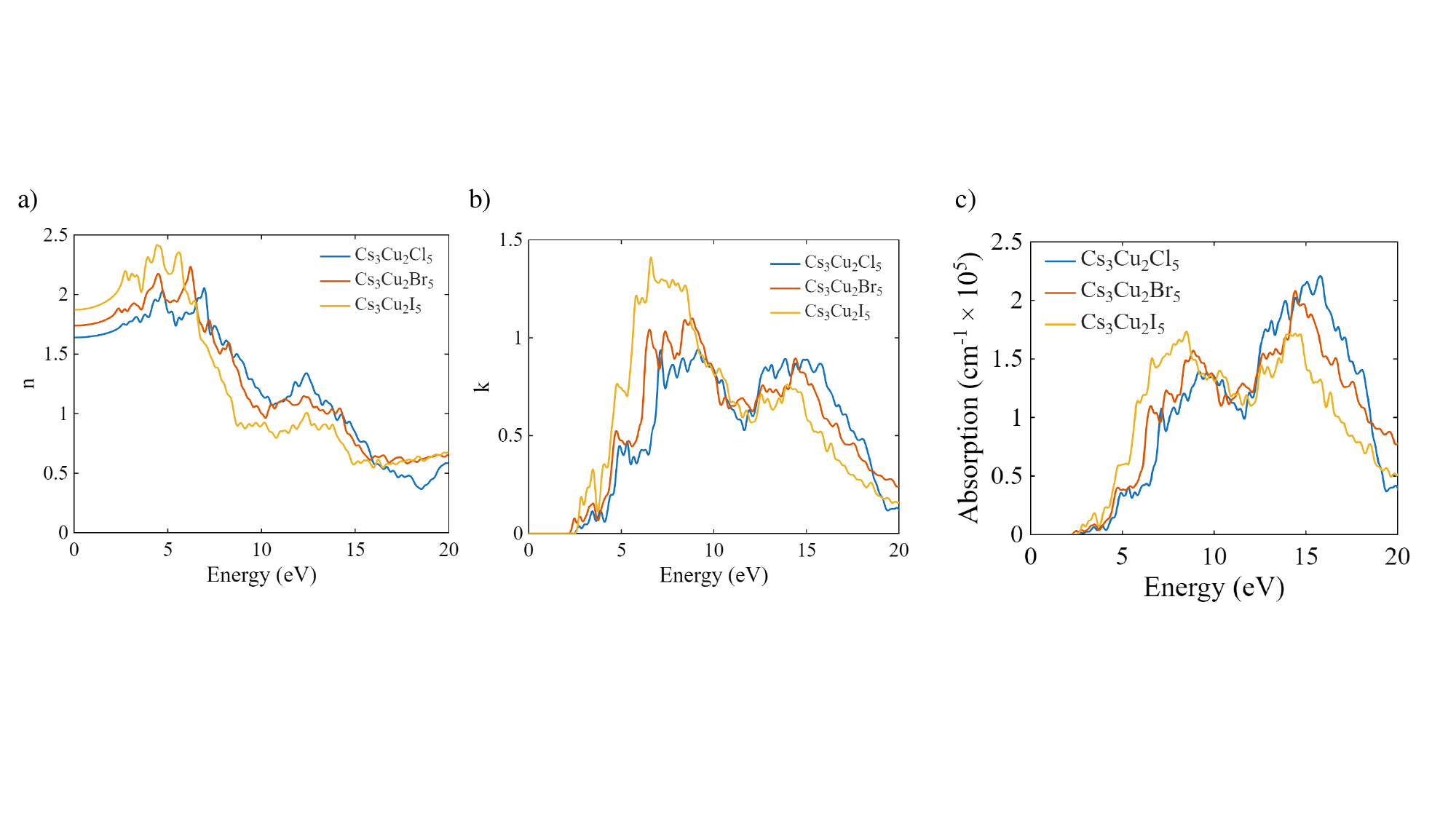}
 \caption{Optical properties of \ce{Cs3Cu2X5} ($X = \ce{Cl,, Br,, I}$): (a) $n$, (b) $k$, and (c) absorption coefficient as a function of photon energy. The calculated spectra are presented in the energy range of 0--20~eV for \ce{Cs3Cu2Cl5}, \ce{Cs3Cu2Br5}, and \ce{Cs3Cu2I5}. The absorption coefficient is expressed in cm$^{-1}$ (scaled by $10^{5}$). The distinct peak structures in $n$, $k$, and absorption originate from interband electronic transitions, reflecting the influence of halogen substitution on the optical response of the compounds.}
 \label{fig:n_k}
\end{figure}

\begin{table}[h]
\centering
\small
  \caption{Lattice parameters in the $a$, $b$, and $c$ directions and band gaps $E_{\mathrm{g}}$ of \ce{Cs3Cu2X5} materials.}
  \label{tbl:lattice}
  \begin{tabular}{lllllll}
    \toprule
    \textbf{Material} & \textbf{a (\AA)} & \textbf{b (\AA)} & \textbf{c (\AA)} & \textbf{$E_{\mathrm{g}}$ (eV)} & \textbf{Remarks} \\
    \midrule
    \ce{Cs3Cu2Cl5} & 9.57 & 10.74 & 13.41 & 2.42 & This work \\
                  & 9.18 & 10.51 & 13.14 & 2.45 & Experimental\cite{xie2020highly} \\
                  & 9.60 & 10.79 & 13.59 & 2.34 & DFT\cite{ali2023highly} \\
    \ce{Cs3Cu2Br5} & 9.90 & 11.16 & 13.93 & 2.15 & This work \\
                  & 9.58 & 10.98 & 13.63 & 2.25 & Experimental\cite{zheng2022stable} \\
                  & 9.92 & 11.19 & 13.92 & 2.09 & DFT\cite{ali2023highly} \\
    \ce{Cs3Cu2I5}  & 10.46 & 11.89 & 14.70 & 2.34 & This work \\
                  & 10.22 & 11.66 & 14.40 & 2.44 & Experimental \\
                  & 10.50 & 11.89 & 14.78 & 2.29 & DFT\cite{ali2023highly} \\
    \bottomrule
  \end{tabular}
\end{table}

Figure~\ref{fig:dos} shows the computed band structures and total density of states for all three \ce{Cs3Cu2X5} compounds where X is Cl, Br, or I. We plotted band dispersions along high-symmetry paths in the Brillouin zone and set the Fermi energy to zero as a reference point. Each of the three materials behaves as a semiconductor with a direct band gap, meaning the valence band maximum and conduction band minimum occur at the same k-point.

For Cs$_3$Cu$_2$Cl$_5$ (Fig.~\ref{fig:dos}a), a direct band gap of 2.42 eV is obtained. Upon substitution of Cl with Br (Fig.~\ref{fig:dos}b), the band gap decreases to 2.15 eV. In the case of Cs$_3$Cu$_2$I$_5$ (Fig.~\ref{fig:dos}c), the band gap slightly increases to 2.34 eV. The direct nature of the band gap is preserved across the halide series, indicating that both the conduction band minimum (CBM) and valence band maximum (VBM) occur at the same high-symmetry $k$-point.

The valence bands are flat near the VBM, which points to localized states from Cu--$X$ orbital mixing. The conduction bands have more dispersion, so carriers should have more mobility in this region. The total DOS plots (right panels of Fig.~\ref{fig:dos}) show clean band gaps with no mid-gap states.

Table~1 lists the optimized lattice parameters and band gaps. Calculated values for \ce{Cs3Cu2Cl5} ($a = 9.57$~\AA, $b = 10.74$~\AA, $c = 13.41$~\AA, $E_g = 2.42$~eV) match well with experiments and earlier experimental and theoretical works. \ce{Cs3Cu2Br5} comes out at $a = 9.90$~\AA, $b = 11.16$~\AA, $c = 13.93$~\AA\ with a 2.15~eV gap. \ce{Cs3Cu2I5} has $a = 10.46$~\AA, $b = 11.89$~\AA, $c = 14.70$~\AA\ and a 2.34~eV gap, again matching the literature.

The lattice expands from Cl to Br to I as the halide ions get larger in size. Longer Cu--$X$ bonds change the orbital overlap and shift the band gaps across the series. Overall, the structural and electronic properties obtained in this work demonstrate strong consistency with experimental and previously reported DFT results, confirming the reliability of the computational methodology employed.
The optical absorption coefficient is fundamentally linked to the complex dielectric function, expressed as $\epsilon = \epsilon_1 + i\epsilon_2$. The imaginary component, $\epsilon_2$, can be determined by examining interband transitions within the electronic band structure, yielding the following relation~\cite{gajdovs2006linear}:
\begin{equation}
\epsilon_2\left(q \rightarrow 0_{\hat{u}}, \hbar\omega\right) = \frac{2e^2\pi}{\Omega\epsilon_0} \sum_{k,v,c} \left|\langle\Psi_k^c|\mathbf{u} \cdot \mathbf{r}|\Psi_k^v\rangle\right|^2 \delta\left(E_k^c - E_k^v - \hbar\omega\right)
\label{eq:epsilon2}
\end{equation}
In this expression, $\mathbf{u}$ represents the polarization vector of the incident light, $e$ is the elementary charge, and $\hbar$ denotes the reduced Planck constant. The term $\langle\Psi_k^c|\mathbf{u} \cdot \mathbf{r}|\Psi_k^v\rangle$ corresponds to the transition matrix element utilizing the momentum operator $\mathbf{u} \cdot \mathbf{r}$. Furthermore, $E_k^c$ and $E_k^v$ designate the energy levels of the conduction and valence bands, respectively, at a given wave vector $k$, while $\omega$ is the angular frequency.

To evaluate the real component of the dielectric function, $\epsilon_1$, the Kramers--Kronig relation is applied~\cite{gajdovs2006linear}:
\begin{equation}
\epsilon_1(\omega) = 1 + \frac{2}{\pi} P \int_0^{\infty} \frac{\epsilon_2(\omega^*)\omega^*}{\omega^{*2} - \omega^2} d\omega^*
\label{eq:epsilon1}
\end{equation}
where $P$ indicates the Cauchy principal value of the integral. 

Based on these dielectric components, the material's optical absorption coefficient, $\alpha$, can be computed using the following equation~\cite{saha2025investigating}:
\begin{equation}
\alpha(\omega) = \frac{4\pi\kappa(\omega)}{\lambda\sqrt{2}} \left(\sqrt{\epsilon_1^2(\omega) + \epsilon_2^2(\omega)} - \epsilon_1(\omega)\right)^{1/2}
\label{eq:absorption}
\end{equation}
where $\lambda$ represents the wavelength of the light and $\kappa$ stands for the extinction coefficient. 

Finally, the frequency-dependent refractive index, $n(\omega)$, and extinction coefficient, $k(\omega)$, are directly derived from the real and imaginary parts of the dielectric function via the following expressions\cite{saha2025investigating}:
\begin{equation}
n(\omega) = \frac{1}{\sqrt{2}} \left[\sqrt{\epsilon_1^2(\omega) + \epsilon_2^2(\omega)} + \epsilon_1(\omega)\right]^{1/2}
\label{eq:n}
\end{equation}
\begin{equation}
k(\omega) = \frac{1}{\sqrt{2}} \left[\sqrt{\epsilon_1^2(\omega) + \epsilon_2^2(\omega)} - \epsilon_1(\omega)\right]^{1/2}
\label{eq:k}
\end{equation}

The calculated optical properties of Cs$_3$Cu$_2$X$_5$ (X = Cl, Br, I) are illustrated in Fig.~\ref{fig:n_k}, including the real part of the refractive index ($n$), extinction coefficient ($k$), and absorption coefficient as functions of photon energy in the range of 0--20~eV.

Figure~\ref{fig:n_k}(a) presents the real part of the refractive index ($n$). In the low-energy region, all compounds exhibit relatively high static refractive indices, indicating strong polarization response. Prominent peaks in $n$ are observed in the ultraviolet (UV) region, which originate from interband electronic transitions. Among the three compounds, noticeable variations in peak positions and magnitudes are observed, reflecting the influence of halogen substitution on the electronic polarizability and band structure.

The extinction coefficient ($k$), shown in Fig.~\ref{fig:n_k}(b), remains nearly zero in the low-energy region below the band gap, confirming the semiconducting nature of these materials. As the photon energy exceeds the band gap, $k$ increases sharply due to the onset of interband transitions. Distinct peaks appear in the UV region, corresponding to strong optical transitions between the valence and conduction bands. The halogen-dependent shift in peak positions further demonstrates the tunability of optical response through chemical substitution.

Figure~\ref{fig:n_k}(c) shows how absorption varies with photon energy. The absorption edge lines up with the band gap for each compound, as expected from the electronic structure. All three absorb strongly in the UV with peaks around $10^{5}$~cm$^{-1}$. The differences between \ce{Cs3Cu2Cl5}, \ce{Cs3Cu2Br5}, and \ce{Cs3Cu2I5} come from how the orbitals mix and how the bands disperse when you change the halogen.

The optical spectra confirm strong UV transitions across the series. Changing the halide tunes the optical response. These characteristics suggest potential applications in ultraviolet optoelectronic and photonic devices.

\subsection{Composition-Dependent Plasmonic Enhancement}

Table~\ref{tbl:results} lists the optimized nanostructure dimensions and FDTD results for all three halides. \ce{Cs3Cu2Cl5} has the highest Purcell factor (4.4$\times$) and LEE (30\%), while \ce{Cs3Cu2Br5} gives the best spectral overlap ($J_{\text{cos}}$ = 0.955). \ce{Cs3Cu2I5} falls behind on both metrics. The differences come down to how the optical constants of each material interact with the plasmon resonance. Lower refractive index helps the chloride, while the bromide benefits from better wavelength matching with the nanorod LSPR.

\begin{table}[t]
\centering
\scriptsize
  \caption{FDTD results for plasmonic \ce{Cs3Cu2X5}-based PeLEDs.}
  \label{tbl:results}
  \resizebox{\columnwidth}{!}{%
  \begin{tabular}{llllllllll}
    \toprule
    \textbf{X} & \textbf{Composition} & \makecell{\textbf{Emission}\\\textbf{$\lambda$ (nm)}} & \textbf{Nanostructure} & \makecell{\textbf{Radius}\\\textbf{(nm)}} & \makecell{\textbf{Length}\\\textbf{(nm)}} & \textbf{Purcell} & \makecell{\textbf{Peak}\\\textbf{$\lambda$ (nm)}} & \makecell{\textbf{LEE}\\\textbf{(\%)}} & \textbf{$J_{\text{cos}}$} \\
    \midrule
    Cl & \ce{Cs3Cu2Cl5} & 529 & Nanorod & 12 & 50 & 4.4 & 539 & 30 & 0.946 \\
    Br & \ce{Cs3Cu2Br5} & 593 & Nanorod & 18 & 50 & 2.8 & 600 & 26 & 0.955 \\
    I & \ce{Cs3Cu2I5} & 541 & Nanosphere & 15 & --- & 2.4 & 538 & 10 & 0.846 \\
    \bottomrule
  \end{tabular}%
  }
\end{table}

\begin{figure}[t]
    \centering
    \begin{minipage}[b]{0.48\textwidth}
        \centering
        \includegraphics[width=\textwidth]{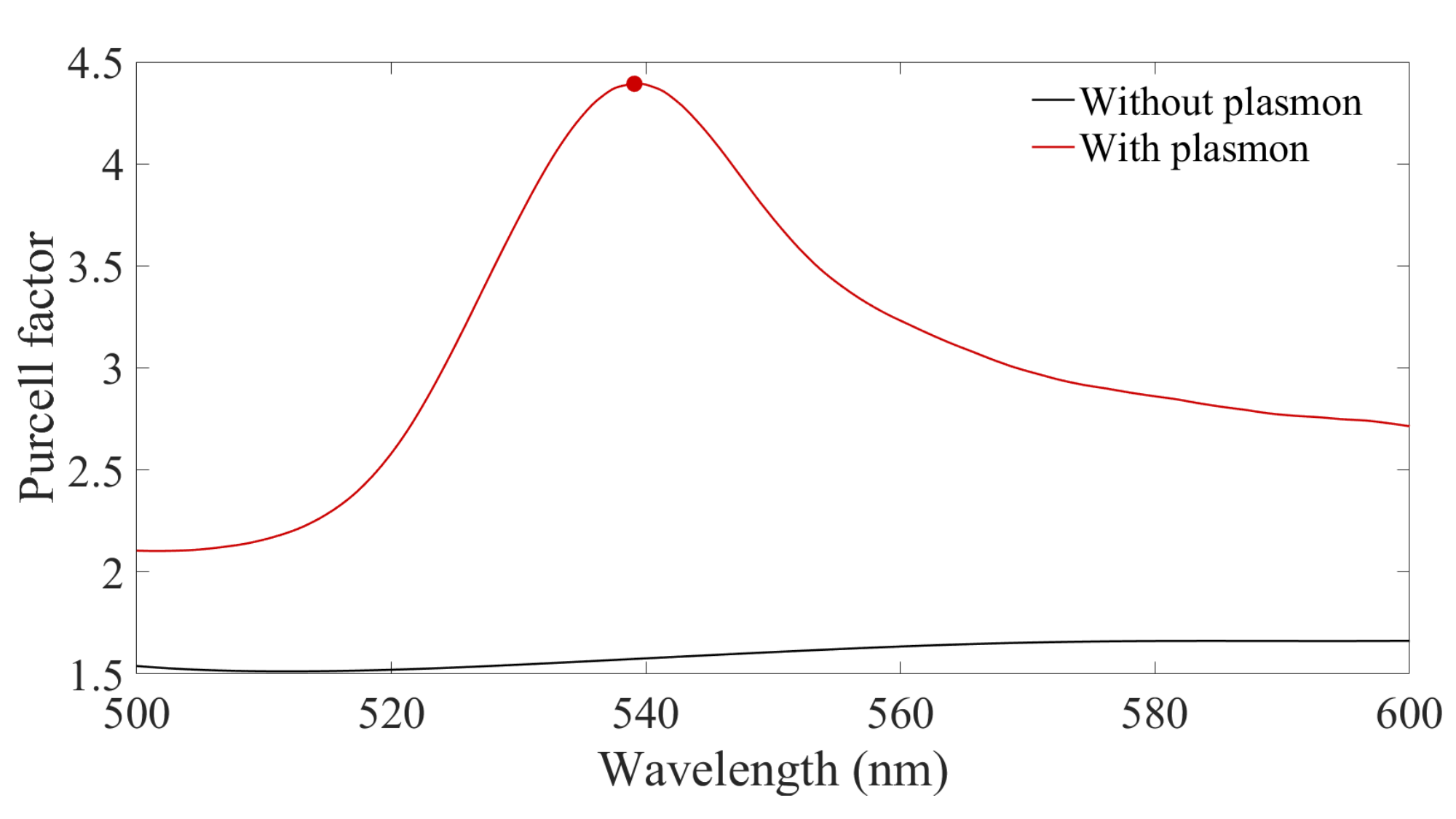}
        (a)
    \end{minipage}
    \hfill
    \begin{minipage}[b]{0.48\textwidth}
        \centering
        \includegraphics[width=\textwidth]{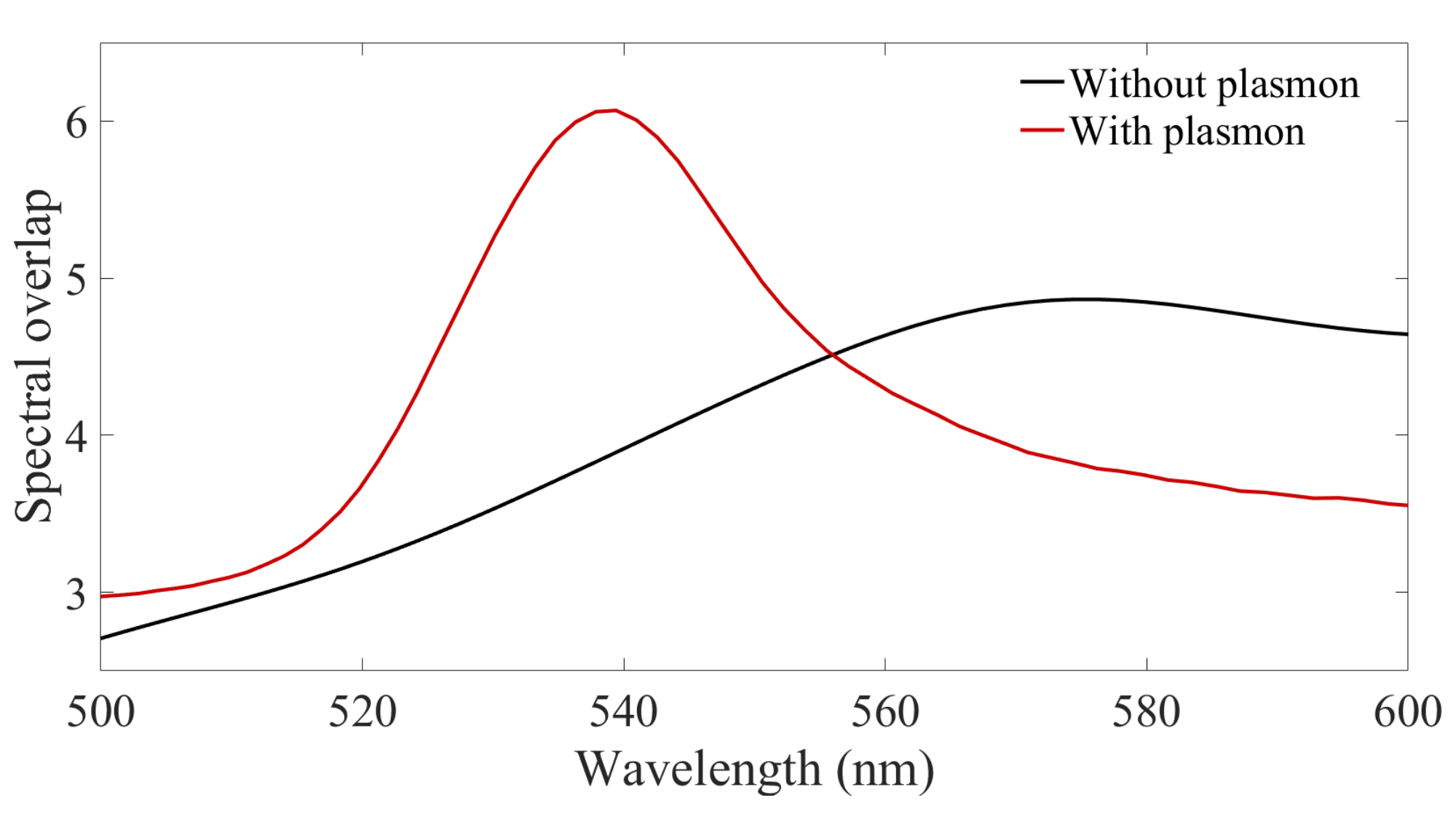}
        (b)
    \end{minipage}
    
    \vspace{0.3cm}
    
    \begin{minipage}[b]{0.48\textwidth}
        \centering
        \includegraphics[width=\textwidth]{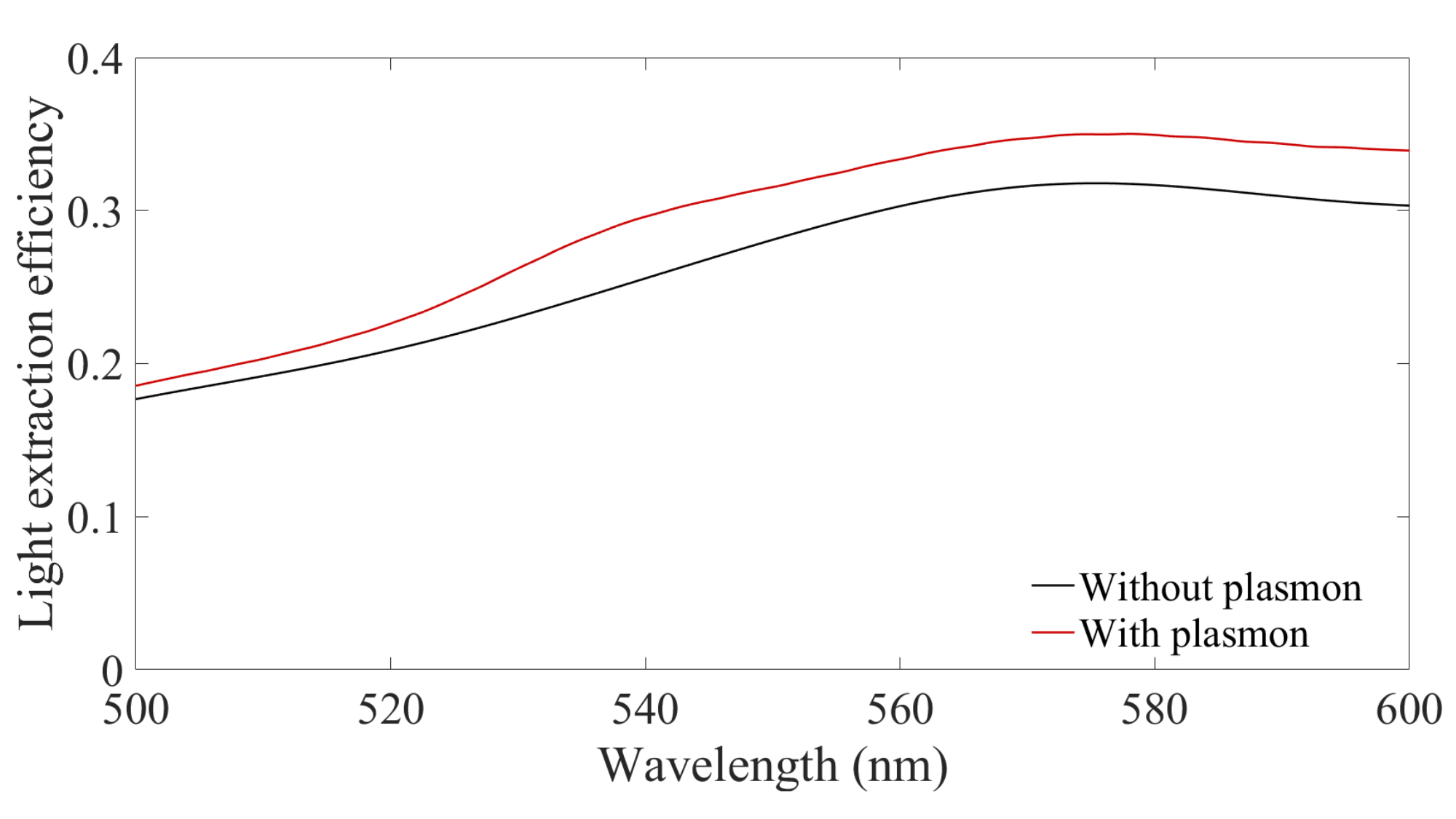}
        (c)
    \end{minipage}
    \hfill
    \begin{minipage}[b]{0.48\textwidth}
        \centering
        \includegraphics[width=\textwidth]{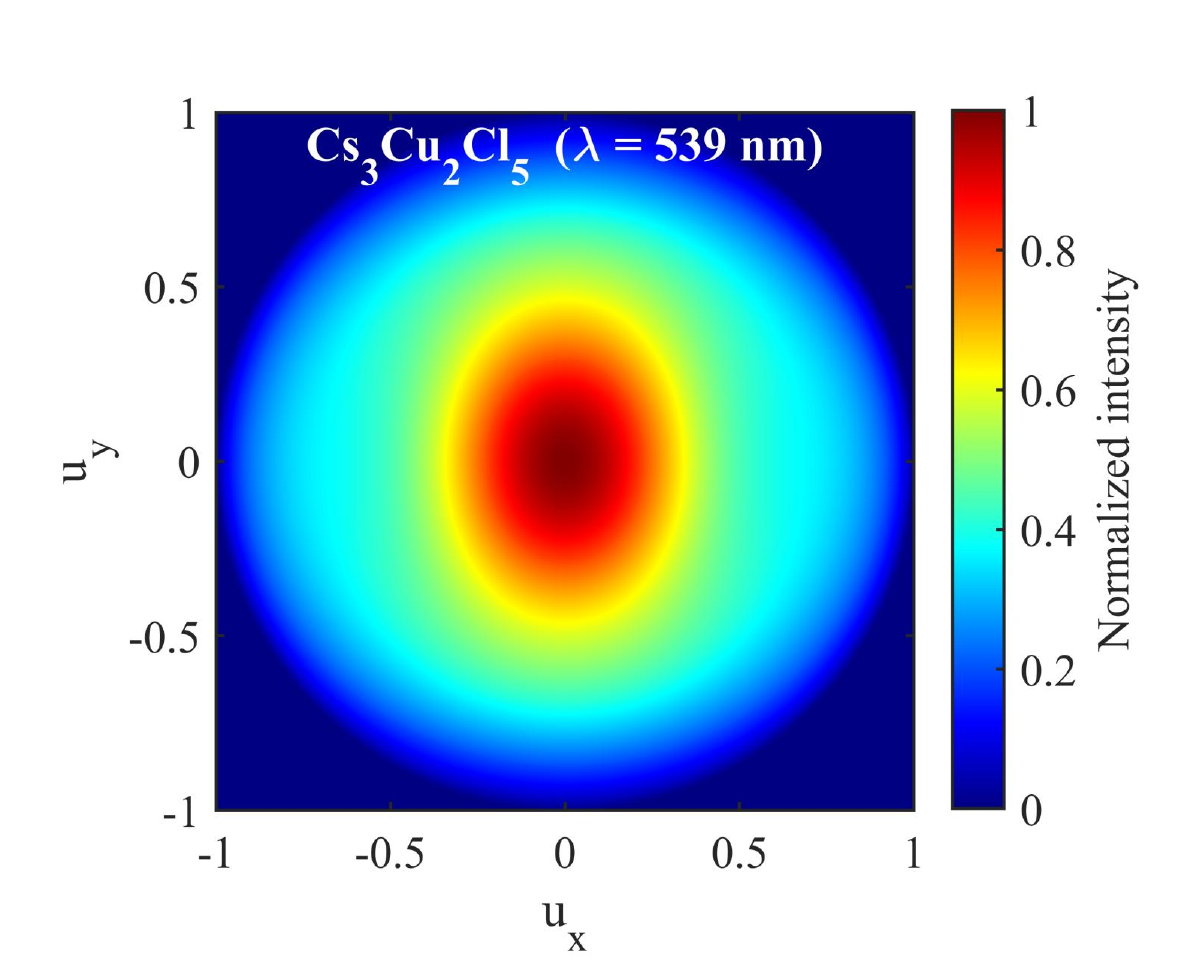}
        (d)
    \end{minipage}
  \caption{Device-level optical performance of the \ce{Cs3Cu2Cl5}-based LED 
  with optimized Ag/\ce{SiO2} nanorod. (a) Purcell factor. (b) Spectral overlap. 
  (c) Light extraction efficiency. (d) Far-field radiation pattern at 539~nm.}
  \label{fig:Cl_results}
\end{figure}

\textbf{\ce{Cs3Cu2Cl5}.} Figure~\ref{fig:Cl_results} shows the device-level 
performance for the \ce{Cs3Cu2Cl5}-based LED. A 4.4$\times$ Purcell enhancement 
is observed at 539~nm (Fig.~\ref{fig:Cl_results}a). This enhancement indicates 
faster spontaneous emission due to increased local density of optical states 
near the Ag nanorod surface. The relatively low refractive index of \ce{Cs3Cu2Cl5} 
($n \approx 1.9$ at 539~nm) reduces dielectric screening of the plasmonic 
near-field. This allows stronger coupling between the emitter dipole and the 
plasmon mode compared to the higher-index Br and I compositions.

The spectral overlap peaks at the same wavelength (Fig.~\ref{fig:Cl_results}b) 
with $J_{\text{cos}} = 0.946$. This confirms good resonance matching between 
the emitter and the nanorod LSPR. Strong spectral alignment is important 
because Purcell enhancement depends on overlap between the emission spectrum 
and plasmonic mode density.

LEE goes from around 18\% without the nanorod to 30\% with it (Fig.~\ref{fig:Cl_results}c), a 1.7$\times$ improvement. The plasmon scatters light that would otherwise remain trapped in waveguide and substrate modes, redirecting it into the far field. The LEE peak appears at slightly longer wavelengths than the Purcell peak. This offset occurs because near-field coupling and far-field scattering do not optimize at the same wavelength. Strong near-field enhancement requires tight mode confinement, while efficient extraction needs good scattering into propagating waves. These two requirements compete to some extent. Regardless, 30\% extraction compares favorably to the 10--20\% typical of perovskite LEDs without optical engineering.

The far-field pattern at 539~nm (Fig.~\ref{fig:Cl_results}d) is smooth and close to Lambertian. Most intensity falls within $\pm$60° of normal, with no secondary lobes. This indicates that the nanorod functions as a scattering center rather than an antenna imposing narrow directionality. Such wide-angle emission is desirable for display applications where uniform brightness across viewing angles matters.

\begin{figure}[t]
    \centering
    \begin{minipage}[b]{0.48\textwidth}
        \centering
        \includegraphics[width=\textwidth]{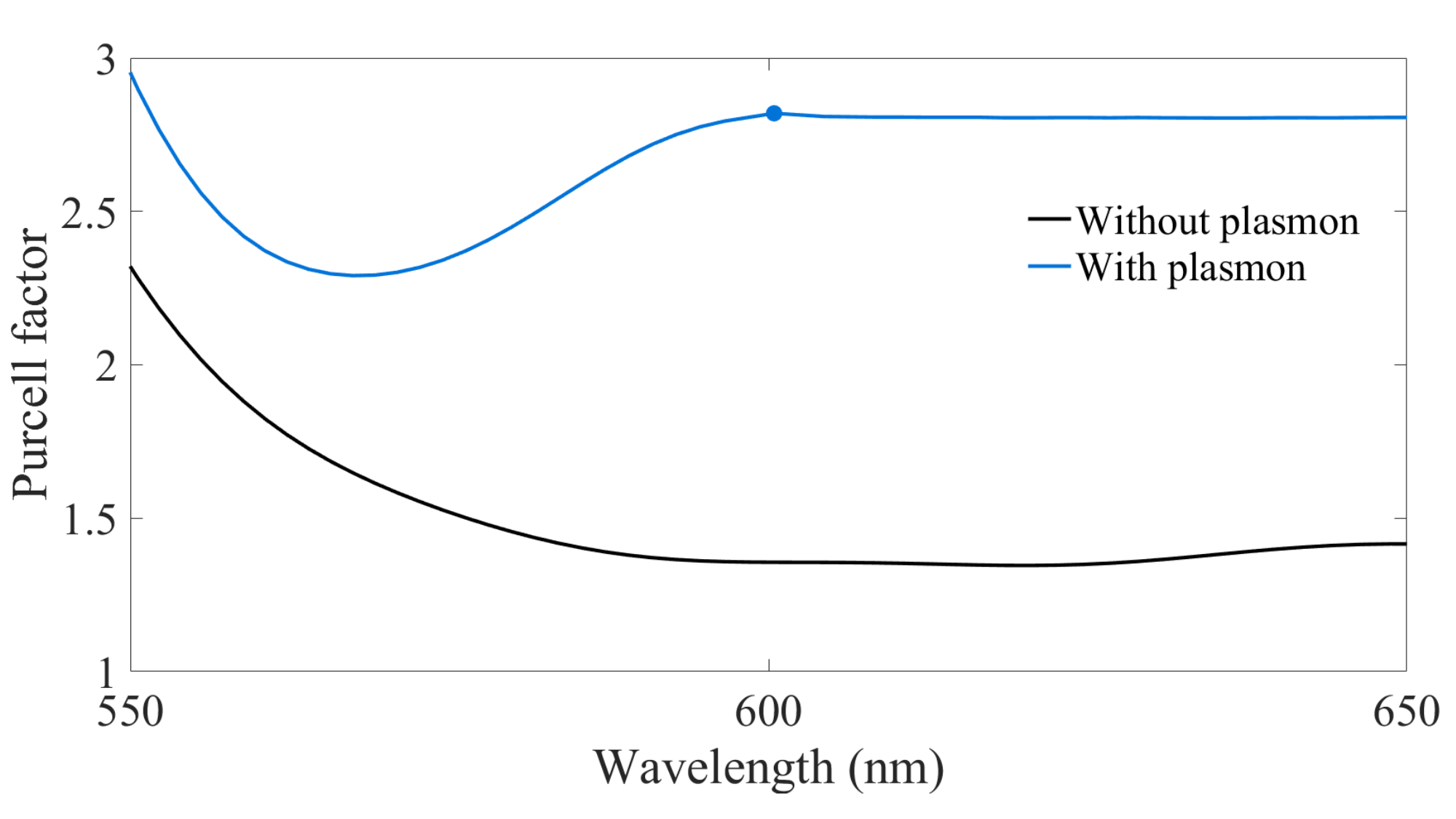}
        (a)
    \end{minipage}
    \hfill
    \begin{minipage}[b]{0.48\textwidth}
        \centering
        \includegraphics[width=\textwidth]{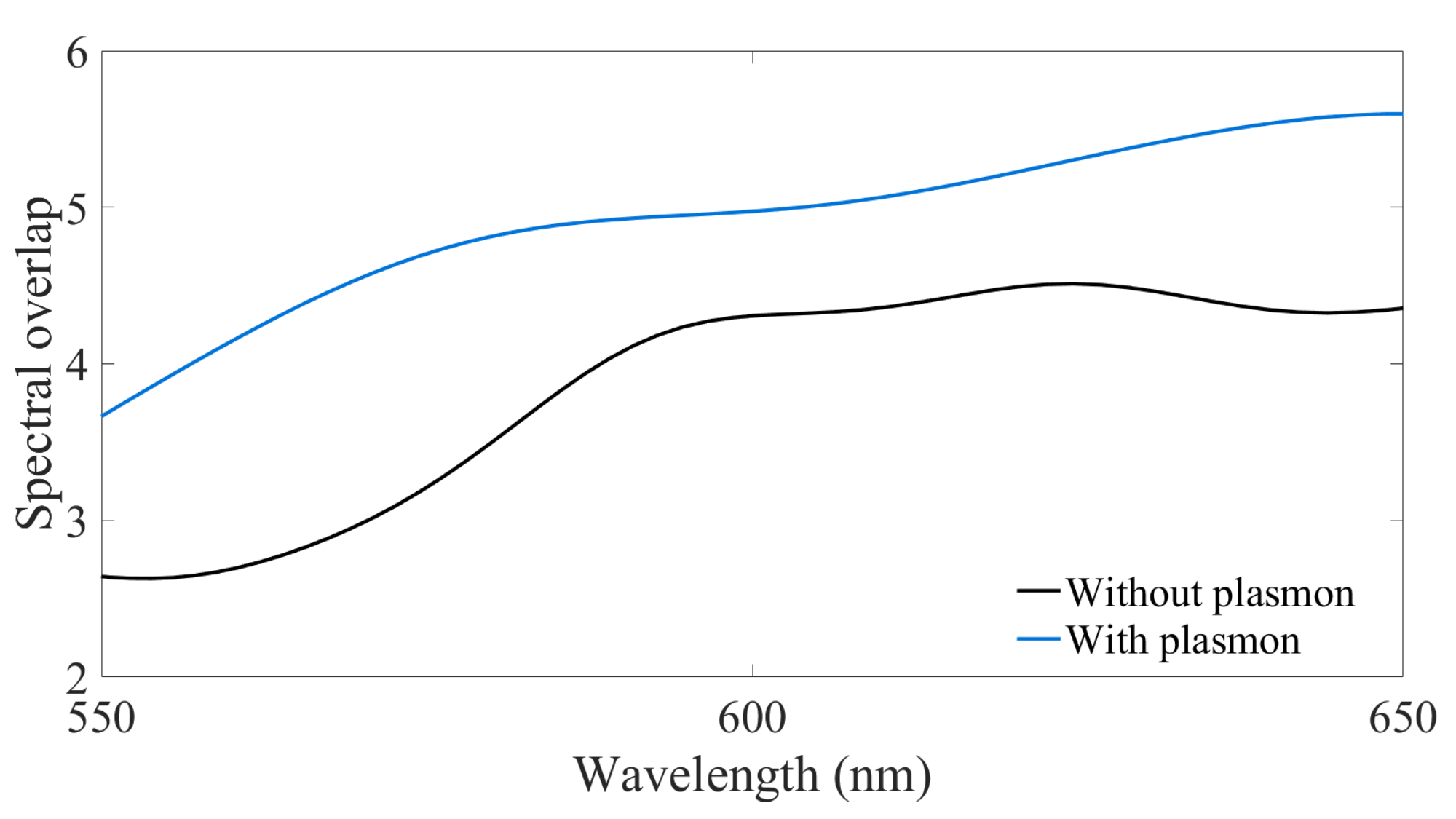}
        (b)
    \end{minipage}
    
    \vspace{0.3cm}
    
    \begin{minipage}[b]{0.48\textwidth}
        \centering
        \includegraphics[width=\textwidth]{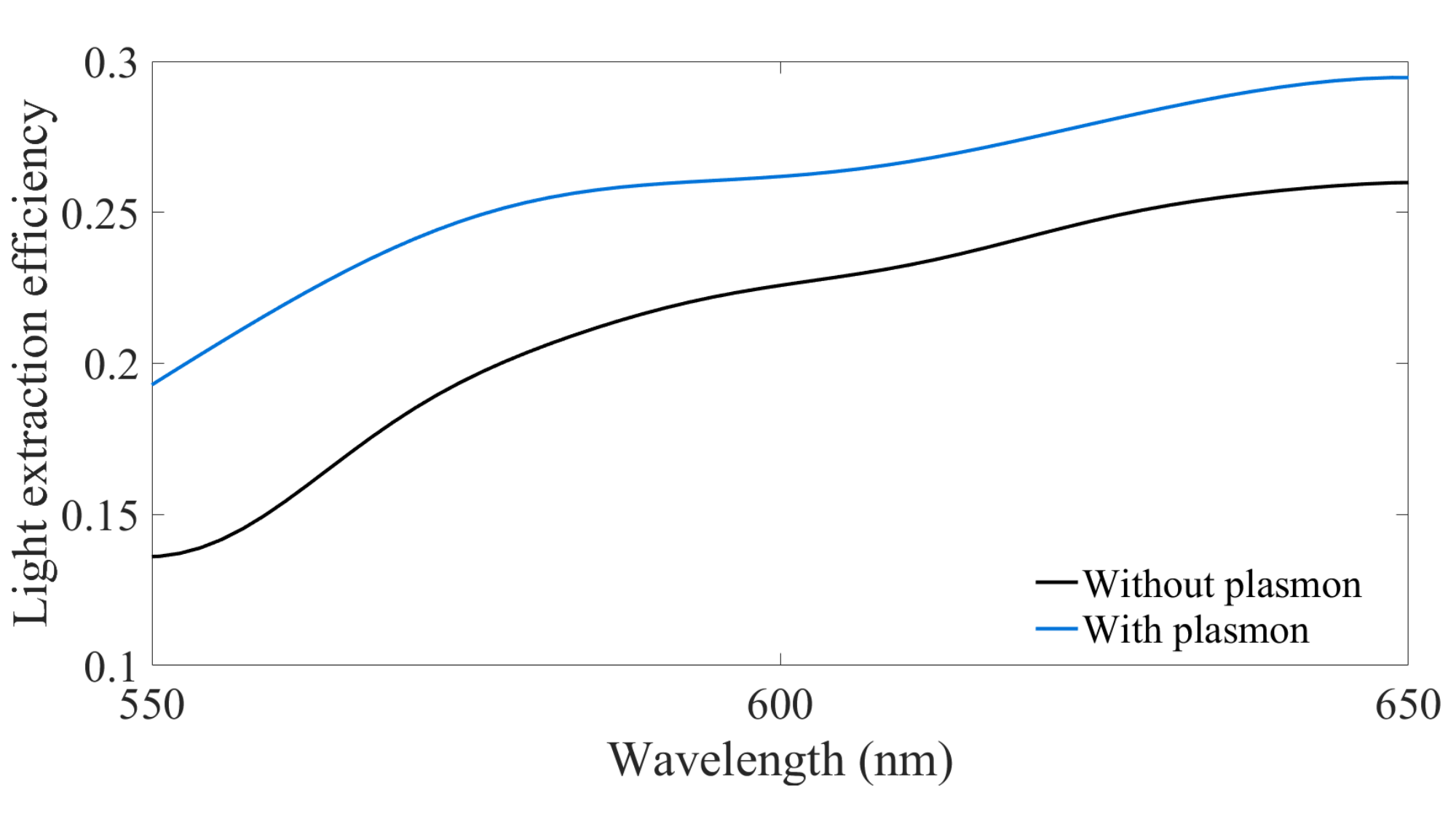}
        (c)
    \end{minipage}
    \hfill
    \begin{minipage}[b]{0.48\textwidth}
        \centering
        \includegraphics[width=\textwidth]{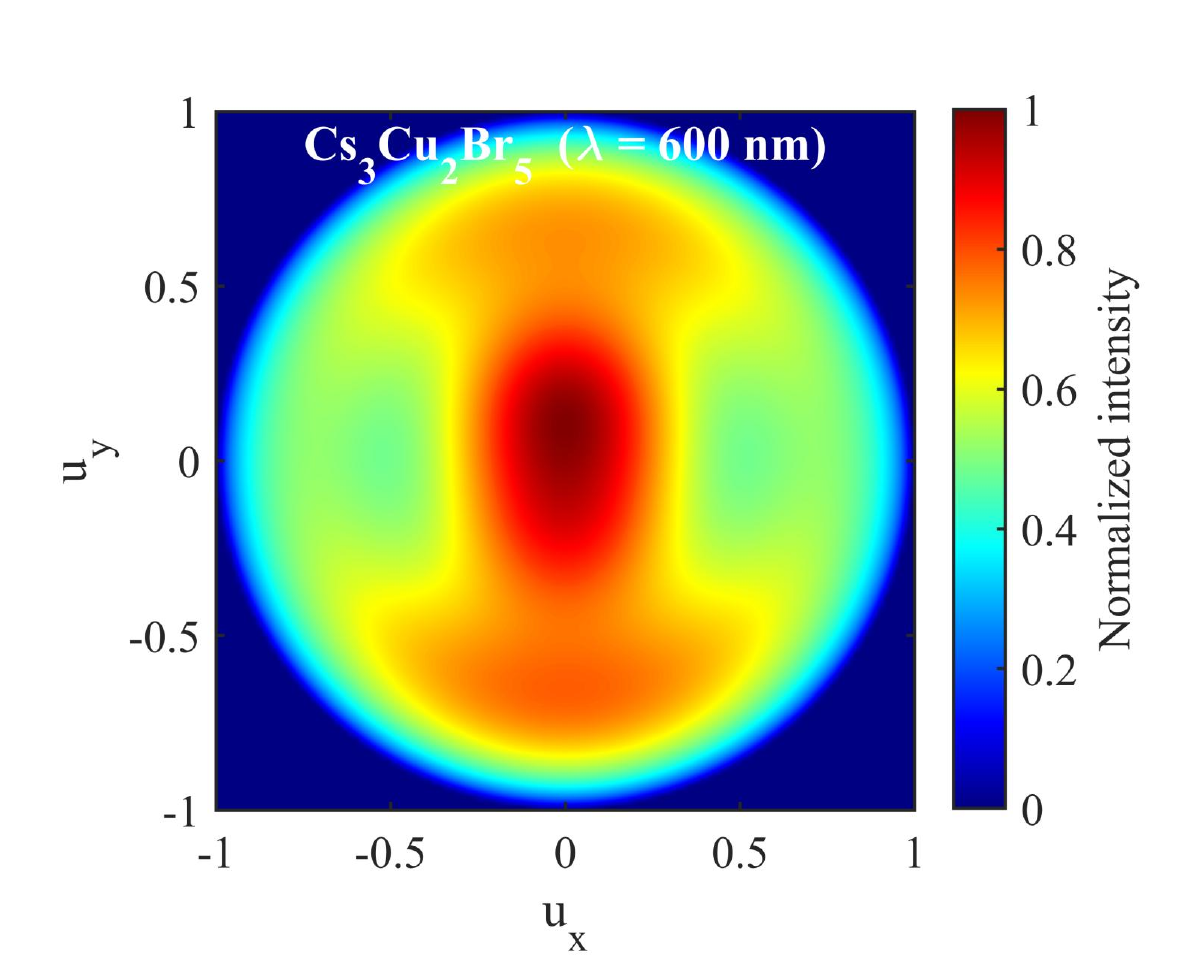}
        (d)
    \end{minipage}
  \caption{Device-level optical performance of the \ce{Cs3Cu2Br5}-based LED 
  with optimized Ag/\ce{SiO2} nanorod. (a) Purcell factor. (b) Spectral overlap. 
  (c) Light extraction efficiency. (d) Far-field radiation pattern at 600~nm.}
  \label{fig:Br_results}
\end{figure}

\textbf{\ce{Cs3Cu2Br5}.} The \ce{Cs3Cu2Br5}-based LED shows 2.8$\times$ Purcell 
enhancement at 600~nm (Fig.~\ref{fig:Br_results}a). This is lower than 
\ce{Cs3Cu2Cl5} even though the spectral overlap is the highest among all 
three compositions ($J_{\text{cos}} = 0.955$, Fig.~\ref{fig:Br_results}b). 
This can be explained by the higher refractive index of \ce{Cs3Cu2Br5} 
($n \approx 2.1$ at 600~nm). A higher index increases dielectric screening 
around the nanorod and reduces the effective near-field intensity at the 
emitter location. The Purcell factor scales roughly as $F_P \propto n^{-3}$, 
which explains the reduced enhancement in higher-index materials.

LEE improves from around 20\% to 26\% (Fig.~\ref{fig:Br_results}c). This 
1.3$\times$ enhancement is more modest than for \ce{Cs3Cu2Cl5}. The higher 
refractive index increases the critical angle for total internal reflection. 
As a result, more photons remain trapped in substrate and waveguide modes. 
The extinction coefficient of \ce{Cs3Cu2Br5} is also higher at the emission 
wavelength (Fig.~\ref{fig:Br_results}b), leading to increased reabsorption losses.

The far-field pattern at 600~nm (Fig.~\ref{fig:Br_results}d) is broader than 
that of \ce{Cs3Cu2Cl5}. More intensity appears at oblique angles. This 
broadening suggests that some emitted light couples to guided modes before 
being scattered out. The longer optical path associated with these guided 
modes also contributes to lower LEE.

\begin{figure}[t]
    \centering
    \begin{minipage}[b]{0.48\textwidth}
        \centering
        \includegraphics[width=\textwidth]{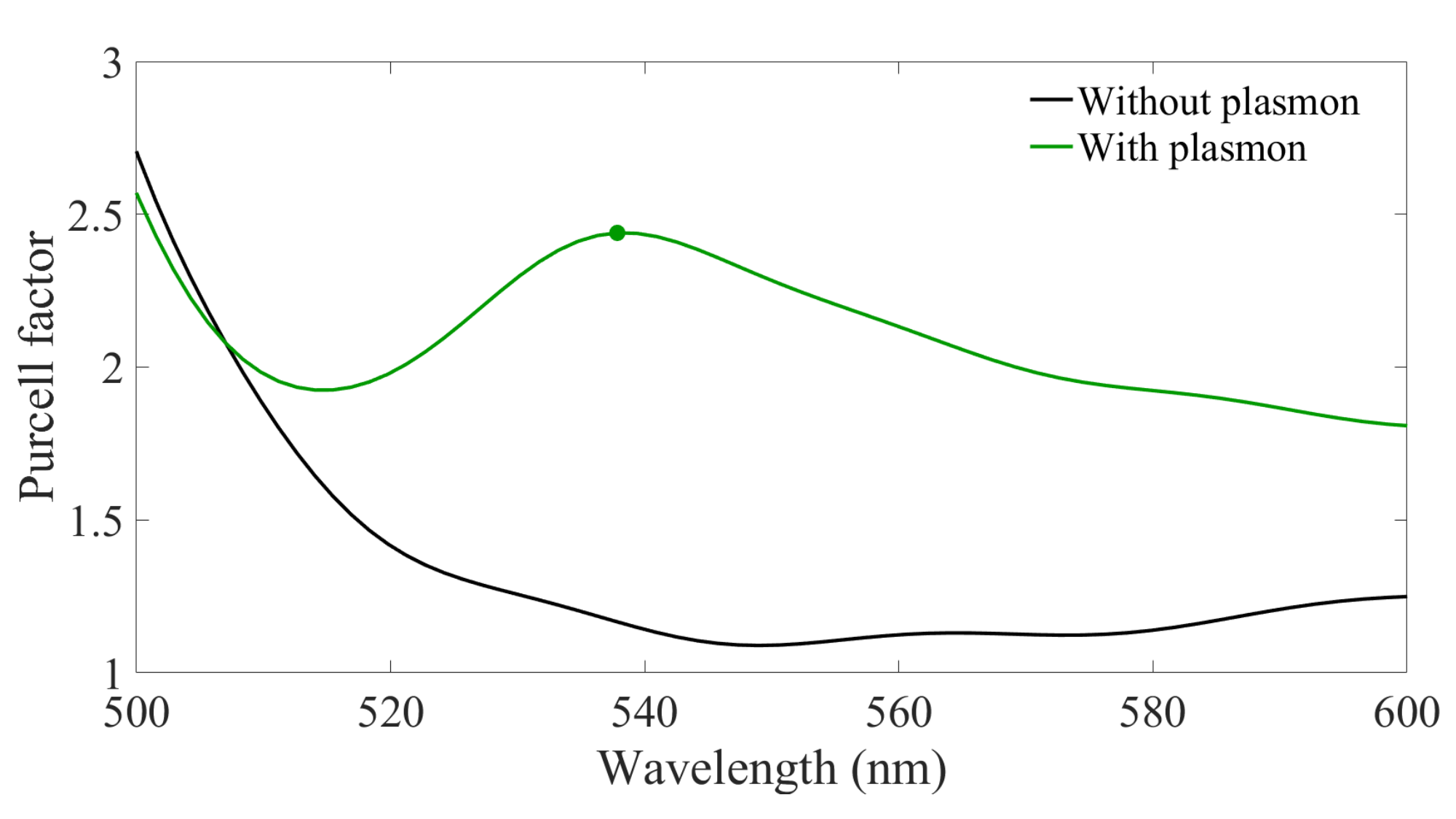}
        (a)
    \end{minipage}
    \hfill
    \begin{minipage}[b]{0.48\textwidth}
        \centering
        \includegraphics[width=\textwidth]{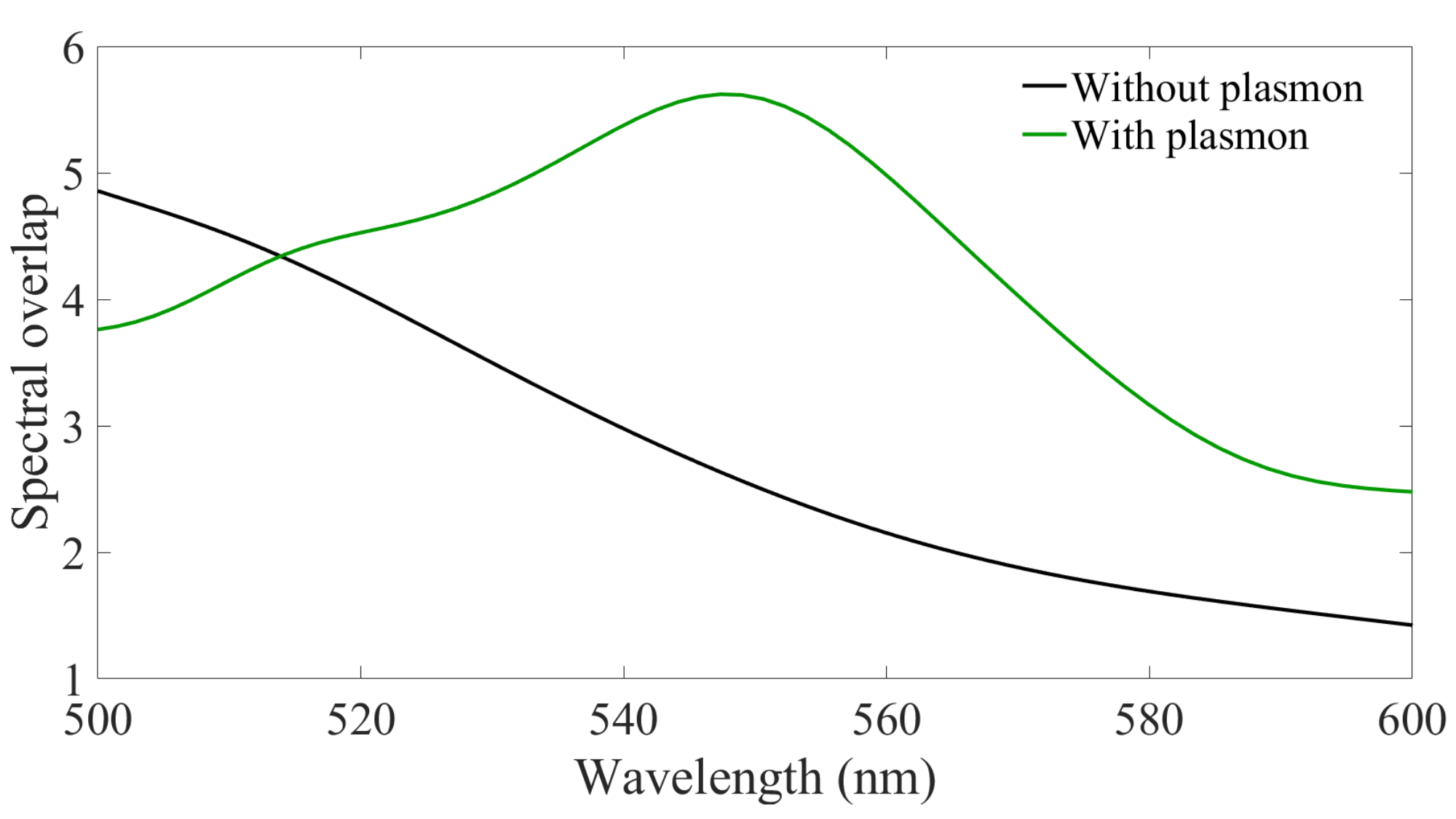}
        (b)
    \end{minipage}
    
    \vspace{0.3cm}
    
    \begin{minipage}[b]{0.48\textwidth}
        \centering
        \includegraphics[width=\textwidth]{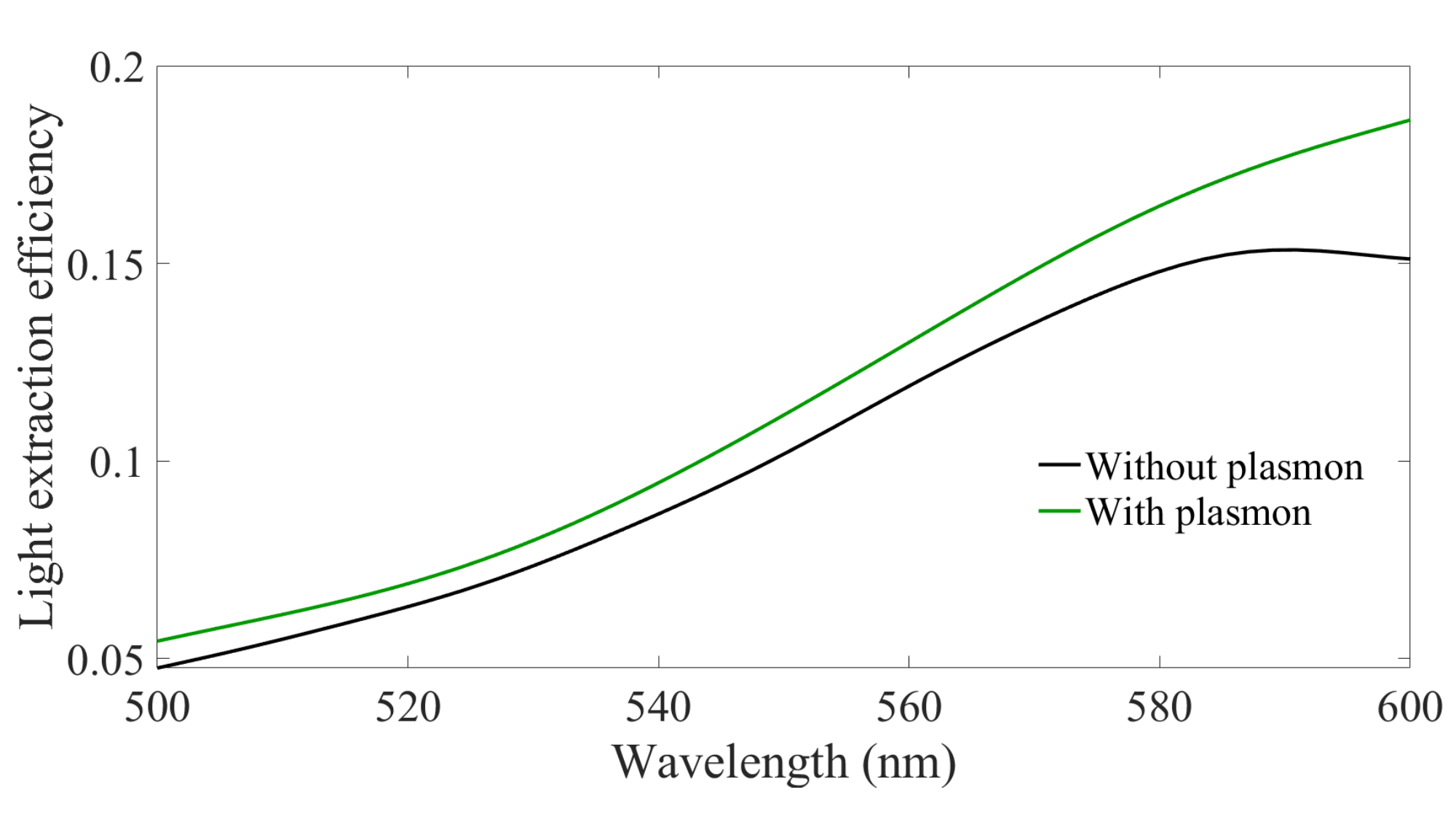}
        (c)
    \end{minipage}
    \hfill
    \begin{minipage}[b]{0.48\textwidth}
        \centering
        \includegraphics[width=\textwidth]{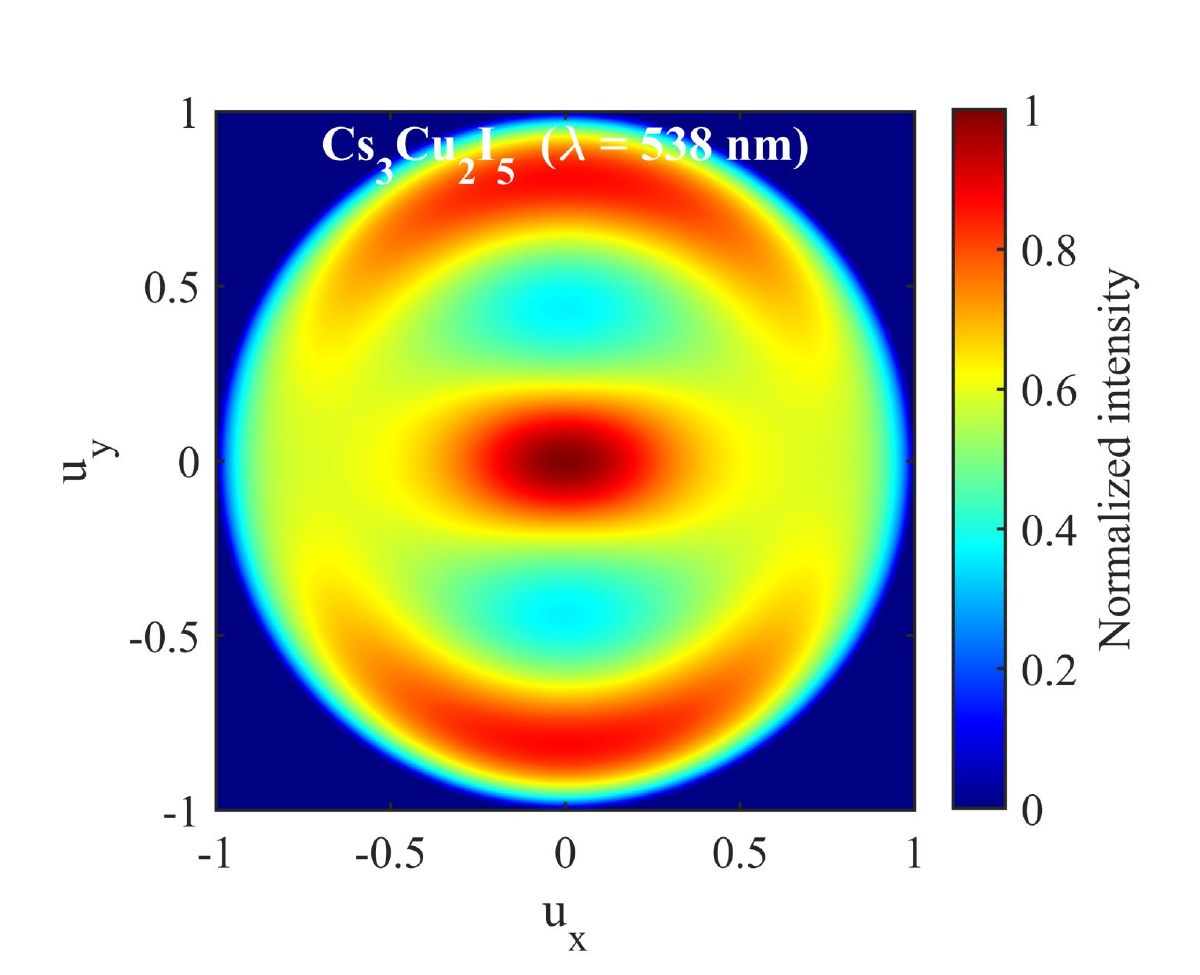}
        (d)
    \end{minipage}
  \caption{Device-level optical performance of the \ce{Cs3Cu2I5}-based LED 
  with optimized Ag/\ce{SiO2} nanosphere. (a) Purcell factor. (b) Spectral 
  overlap. (c) Light extraction efficiency. (d) Far-field radiation pattern 
  at 538~nm.}
  \label{fig:I_results}
\end{figure}

\textbf{\ce{Cs3Cu2I5}.} For \ce{Cs3Cu2I5}, a nanosphere geometry was used 
instead of a nanorod. Parametric optimization showed that nanorods could not 
achieve good spectral overlap with the \ce{Cs3Cu2I5} emission at 541~nm. The 
different dielectric environment created by the iodide composition shifts the 
optimal LSPR condition. The isotropic field distribution of nanospheres 
provides better mode matching with the emission dipole orientation in this 
material.

The Purcell factor peaks at 2.4$\times$ at 538~nm (Fig.~\ref{fig:I_results}a). 
This is the lowest among the three compositions. Without plasmonic coupling, 
the Purcell factor stays near 1.1 to 1.2$\times$. This indicates weak cavity 
effects in the device stack at this emission wavelength. The spectral overlap 
is also the lowest ($J_{\text{cos}} = 0.846$, Fig.~\ref{fig:I_results}b). 
This reflects limited alignment between the nanosphere LSPR and the emission 
spectrum.

LEE remains at only 10\% (Fig.~\ref{fig:I_results}c). This is much lower than 
the Cl and Br compositions. The combination of high refractive index 
($n \approx 2.2$) and substantial extinction coefficient at the emission 
wavelength promotes strong optical confinement and reabsorption. The far-field 
pattern (Fig.~\ref{fig:I_results}d) shows clear distortion with significant 
intensity at high angles. This confirms that many emitted photons couple to 
guided or substrate modes rather than escaping to free space.

The \ce{Cs3Cu2I5} case shows a key limitation. Even with reasonable Purcell enhancement, extraction remains poor. Accelerating emission does not guarantee that photons escape the device. The refractive index and extinction coefficient of the host set an upper bound on extraction, and plasmonic enhancement cannot fully compensate for a high-index environment.

\begin{figure}[htbp]
 \centering
\begin{minipage}[b]{0.7\textwidth}
        \centering
        \includegraphics[width=\textwidth]{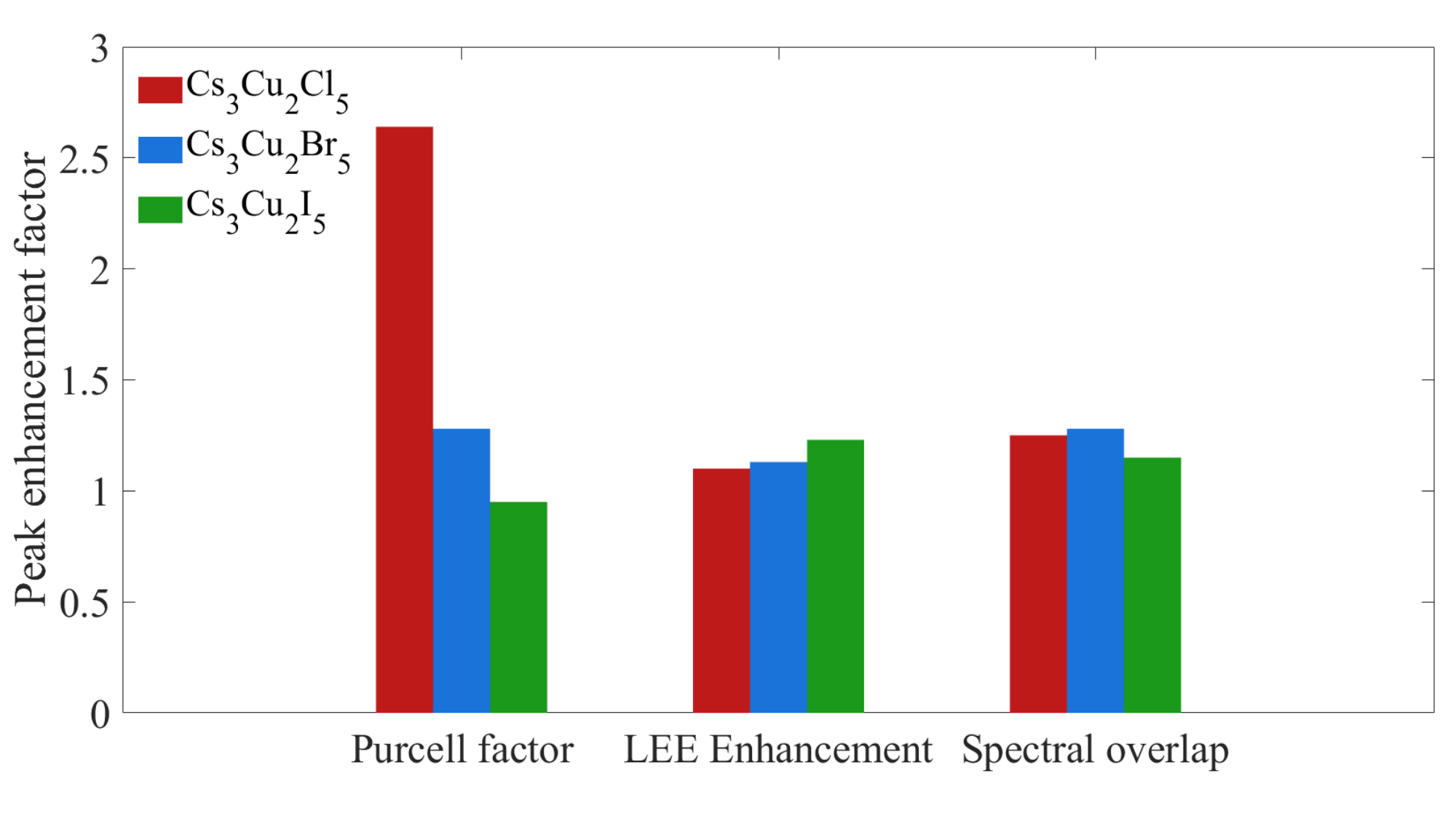}
        (a)
    \end{minipage}
    
    \vspace{0.3cm}
    
    \begin{minipage}[b]{0.7\textwidth}
        \centering
        \includegraphics[width=\textwidth]{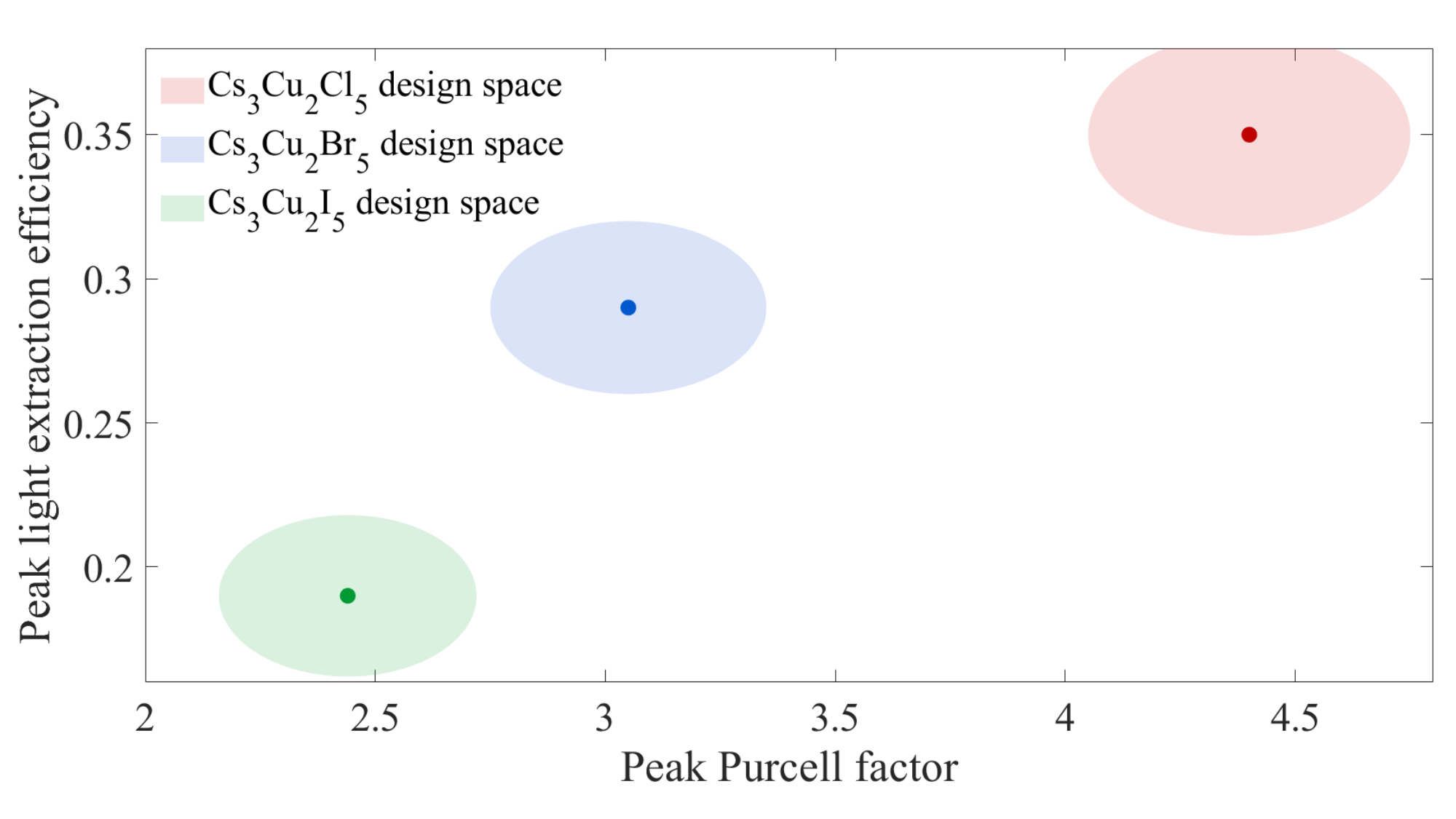}
        (b)
    \end{minipage}
    
    \vspace{0.3cm}
    
    \begin{minipage}[b]{0.7\textwidth}
        \centering
        \includegraphics[width=\textwidth]{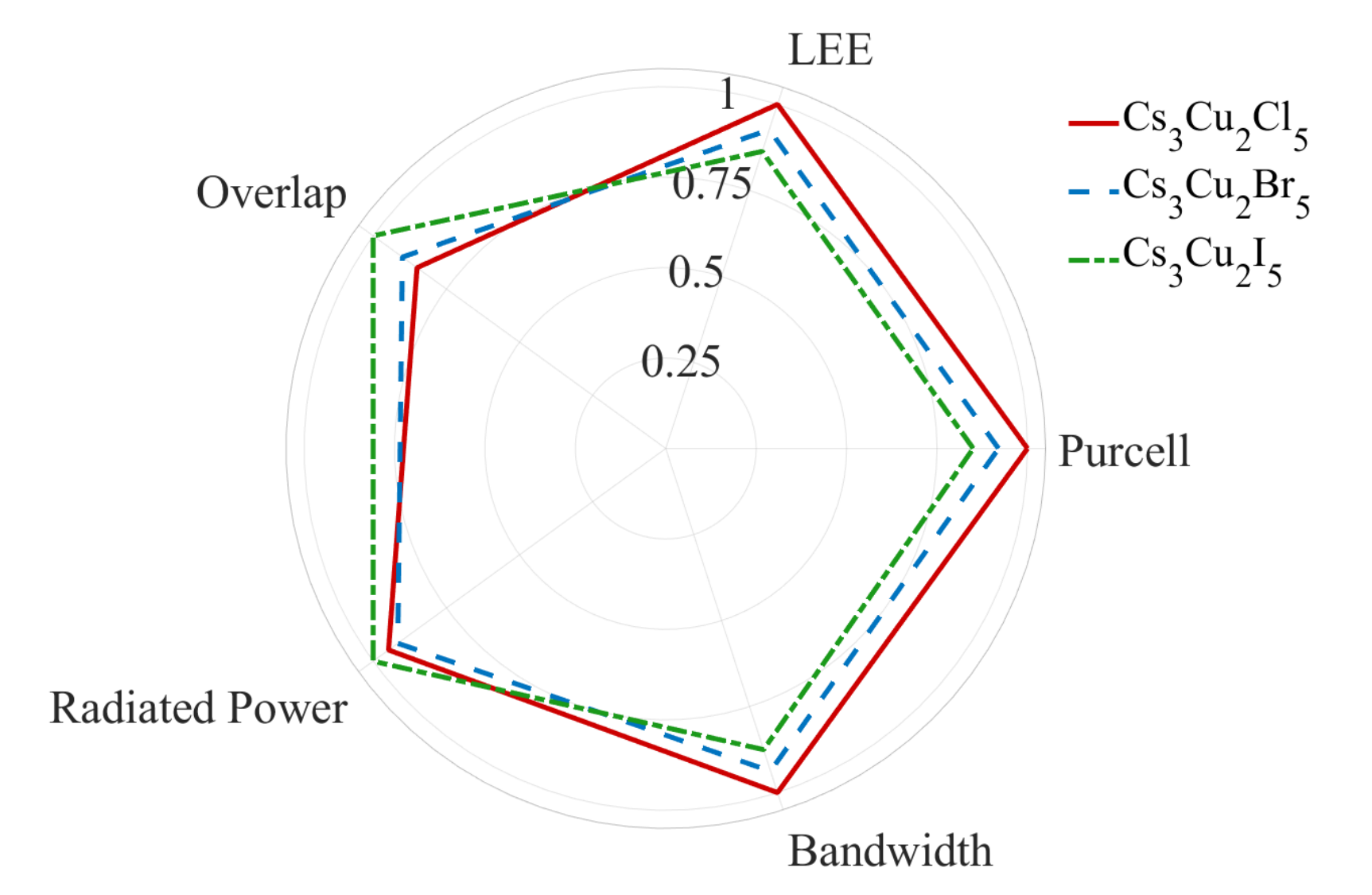}
        (c)
    \end{minipage}
\caption{Comparative performance of plasmon-enhanced \ce{Cs3Cu2X5}-based LEDs. 
(a) Peak enhancement factors for Purcell factor, LEE, and spectral overlap. 
(b) Purcell factor and LEE trade-off map showing the design space for each 
composition. (c) Radar plot comparing normalized performance metrics across 
the halide series.}
 \label{fig:comparison_1}
\end{figure}

\subsection{Comparative Performance and Trade-Off Analysis}

Figure~\ref{fig:comparison_1} puts all three compositions side by side. The differences in plasmonic enhancement across the halide series tie directly to the DFT-derived optical constants, so comparing them together helps explain why some materials respond better than others.

\textbf{Peak Enhancement Comparison.} Figure~\ref{fig:comparison_1}(a) plots the peak enhancement values for each halide. \ce{Cs3Cu2Cl5} leads with a 4.4$\times$ Purcell factor, 30\% LEE, and 0.946 spectral overlap. The chloride does well because its refractive index is low ($n \approx 1.9$), which cuts down dielectric screening and lets the emitter couple more strongly to the nanorod near-field. The plasmonic field penetrates further into the active layer before getting damped. \ce{Cs3Cu2Br5} and \ce{Cs3Cu2I5} have higher indices, so the same nanorod geometry produces weaker enhancement in those hosts.

\ce{Cs3Cu2Br5} exhibits moderate but balanced enhancement across all metrics. 
The higher refractive index ($n \approx 2.1$) limits the Purcell factor to 
2.8$\times$ despite achieving the best spectral overlap (0.955). This shows 
that spectral matching alone cannot compensate for increased dielectric losses 
in higher-index materials.

\ce{Cs3Cu2I5} shows the weakest response with Purcell factor of 2.4$\times$ 
and LEE of only 10\%. The high refractive index ($n \approx 2.2$) and large 
extinction coefficient at the emission wavelength cause strong optical 
confinement. Even with a nanosphere geometry optimized for this composition, 
the spectral overlap remains limited at 0.846.

\textbf{Purcell Factor and LEE Trade-Off.} Figure~\ref{fig:comparison_1}(b) 
maps the relationship between peak Purcell factor and peak LEE. The shaded 
regions represent the accessible design space for each composition through 
nanostructure geometry optimization.

\ce{Cs3Cu2Cl5} occupies the upper-right region where both high Purcell factor 
and high LEE are achievable. This is because the low refractive index creates 
favourable conditions for both near-field enhancement and far-field extraction. 
The plasmonic near-field is less screened, and the critical angle for total 
internal reflection is larger. Both effects work together to improve device 
performance.

\ce{Cs3Cu2Br5} lies in an intermediate region. The higher index material can 
still achieve reasonable Purcell enhancement through geometry optimization. 
However, the increased optical confinement limits how much of this enhanced 
emission can be extracted to the far field.

\ce{Cs3Cu2I5} is confined to the lower-left region. The high refractive index 
creates a fundamental trade-off. Stronger near-field confinement could increase 
the Purcell factor, but this same confinement traps more light inside the 
device. The design space is therefore limited by the intrinsic optical 
properties of the material rather than by the plasmonic structure.

This points to a key design consideration. For plasmon-enhanced LEDs, the choice of emitter material matters as much as the nanostructure geometry. A high Purcell factor in a high-index host may not improve device efficiency if most photons remain trapped.

\textbf{Multi-Parameter Summary.} Figure~\ref{fig:comparison_1}(c) shows a normalized radar plot covering four performance metrics. \ce{Cs3Cu2Cl5} has the largest area, meaning it performs well across Purcell factor, LEE, radiated power, and bandwidth. The balanced shape indicates that plasmonic enhancement and light extraction work together effectively in this material.

\ce{Cs3Cu2Br5} has a smaller but still balanced profile. All metrics drop compared to \ce{Cs3Cu2Cl5}, which follows from the higher refractive index weakening both near-field coupling and far-field extraction.

\ce{Cs3Cu2I5} has the smallest area, with LEE and radiated power particularly low. The shape is also uneven since spectral overlap stays reasonable while extraction suffers. This confirms that optical confinement limits the iodide more than the other two compositions.

Among the \ce{Cs3Cu2X5} series, \ce{Cs3Cu2Cl5} stands out as the best candidate for plasmon-enhanced LEDs. Its low refractive index and strong plasmonic coupling together offer a route to better device efficiency.

\subsection{Dependence of Radiated Power on Wavelength and Emitter-Plasmon Distance}

\begin{figure}[t]
\centering
    \begin{minipage}[b]{0.48\textwidth}
        \centering
        \includegraphics[width=\textwidth]{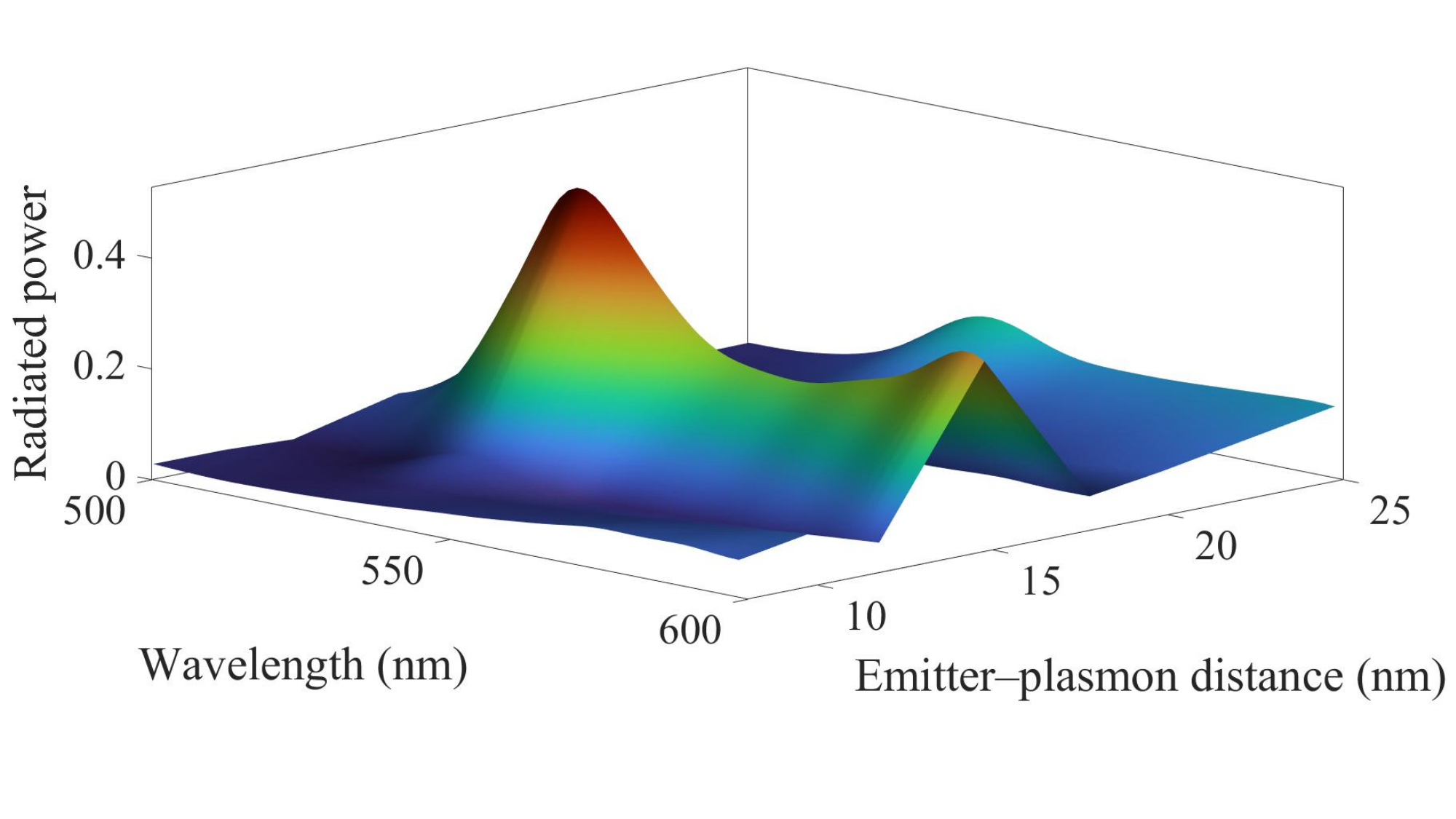}
        (a)
    \end{minipage}
    \hfill
    \begin{minipage}[b]{0.48\textwidth}
        \centering
        \includegraphics[width=\textwidth]{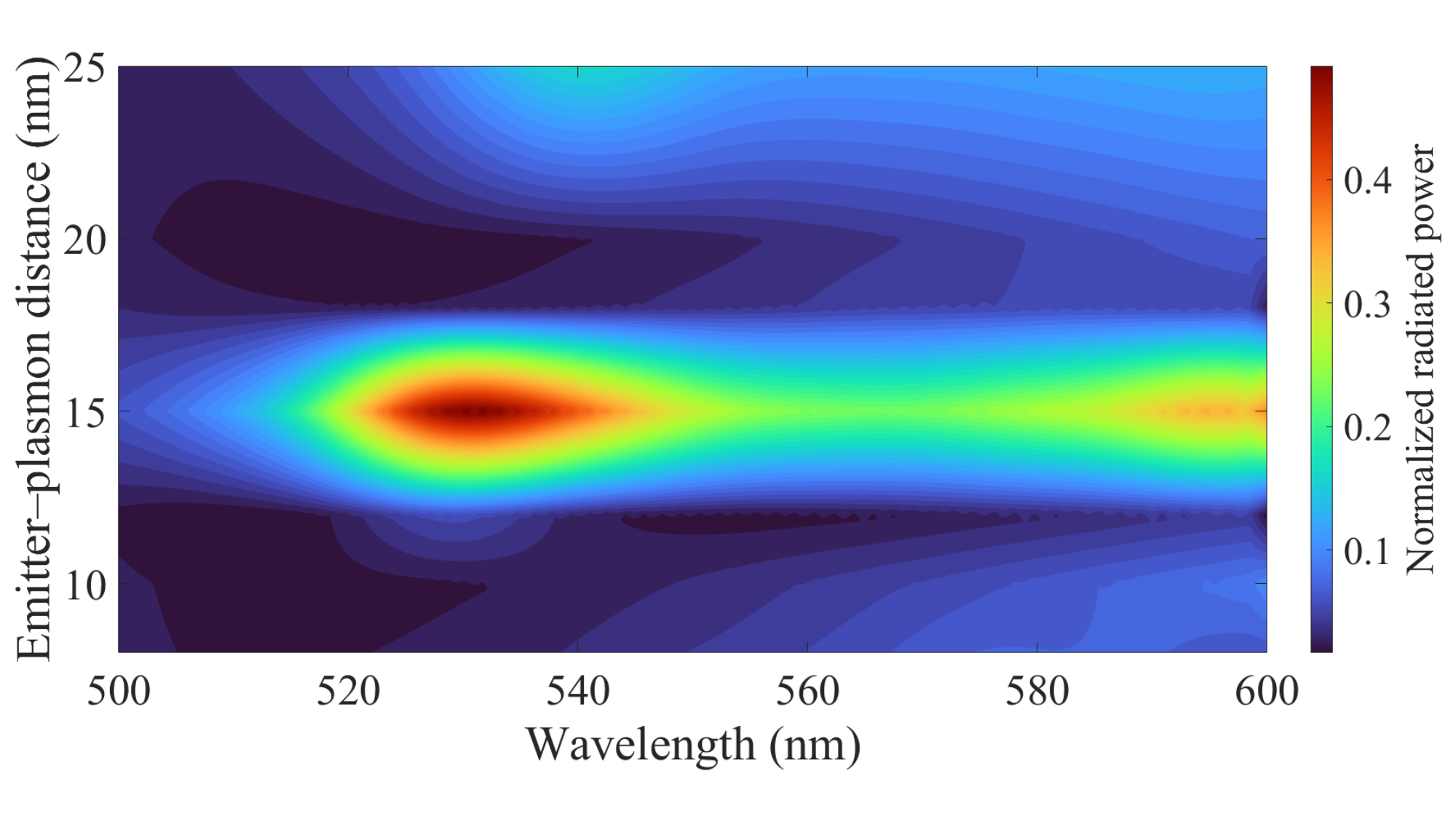}
        (b)
    \end{minipage}
  \caption{Distance and wavelength dependent radiated power for the 
  \ce{Cs3Cu2Cl5}-based LED. (a) 3D surface plot showing a localized maximum 
  at around 15~nm separation and 530 to 540~nm wavelength. (b) 2D contour map 
  showing the narrow optimal coupling window.}
  \label{fig:Cl_radiated}
\end{figure}

\begin{figure}[t]
\centering
    \begin{minipage}[b]{0.48\textwidth}
        \centering
        \includegraphics[width=\textwidth]{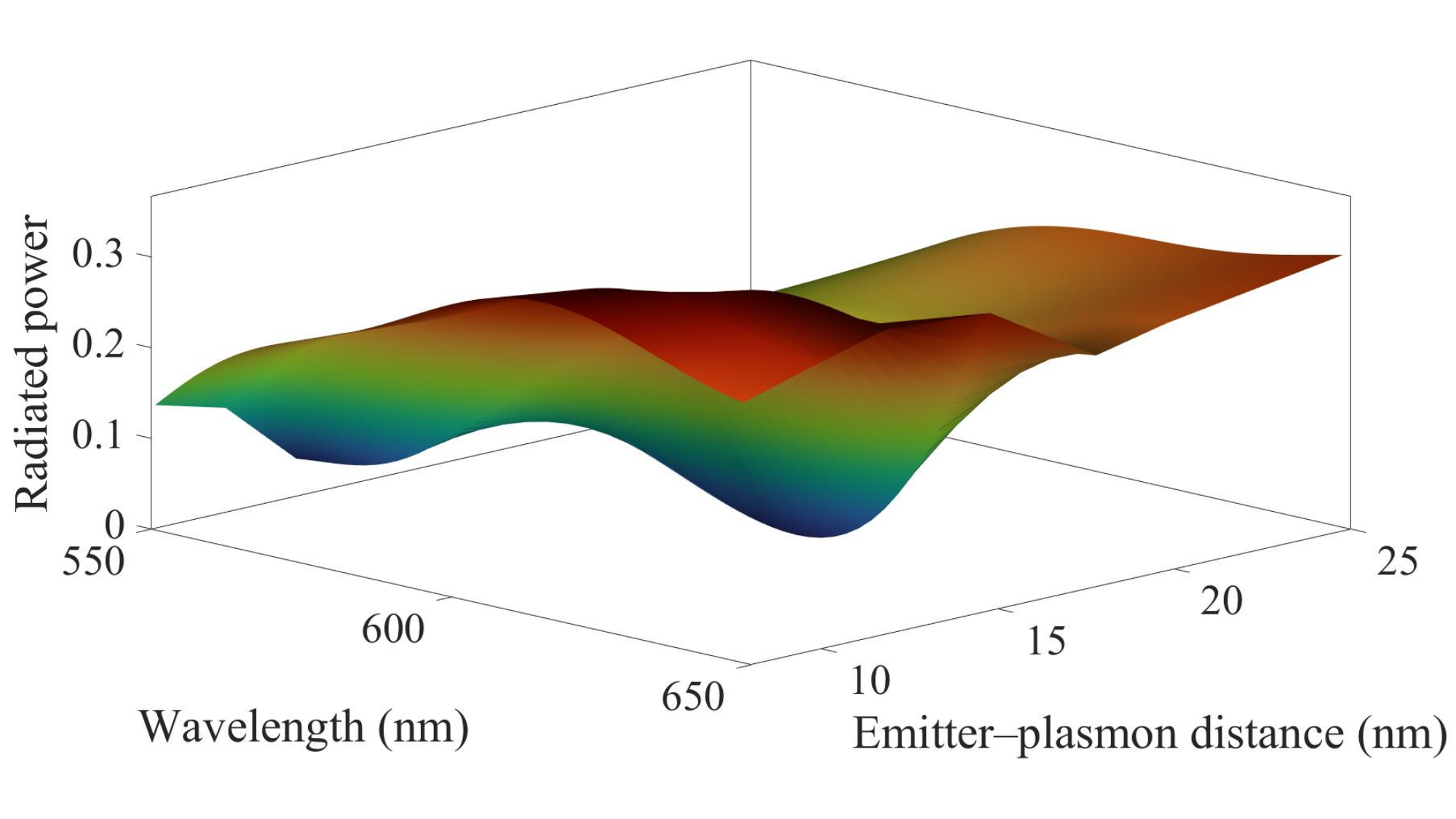}
        (a)
    \end{minipage}
    \hfill
    \begin{minipage}[b]{0.48\textwidth}
        \centering
        \includegraphics[width=\textwidth]{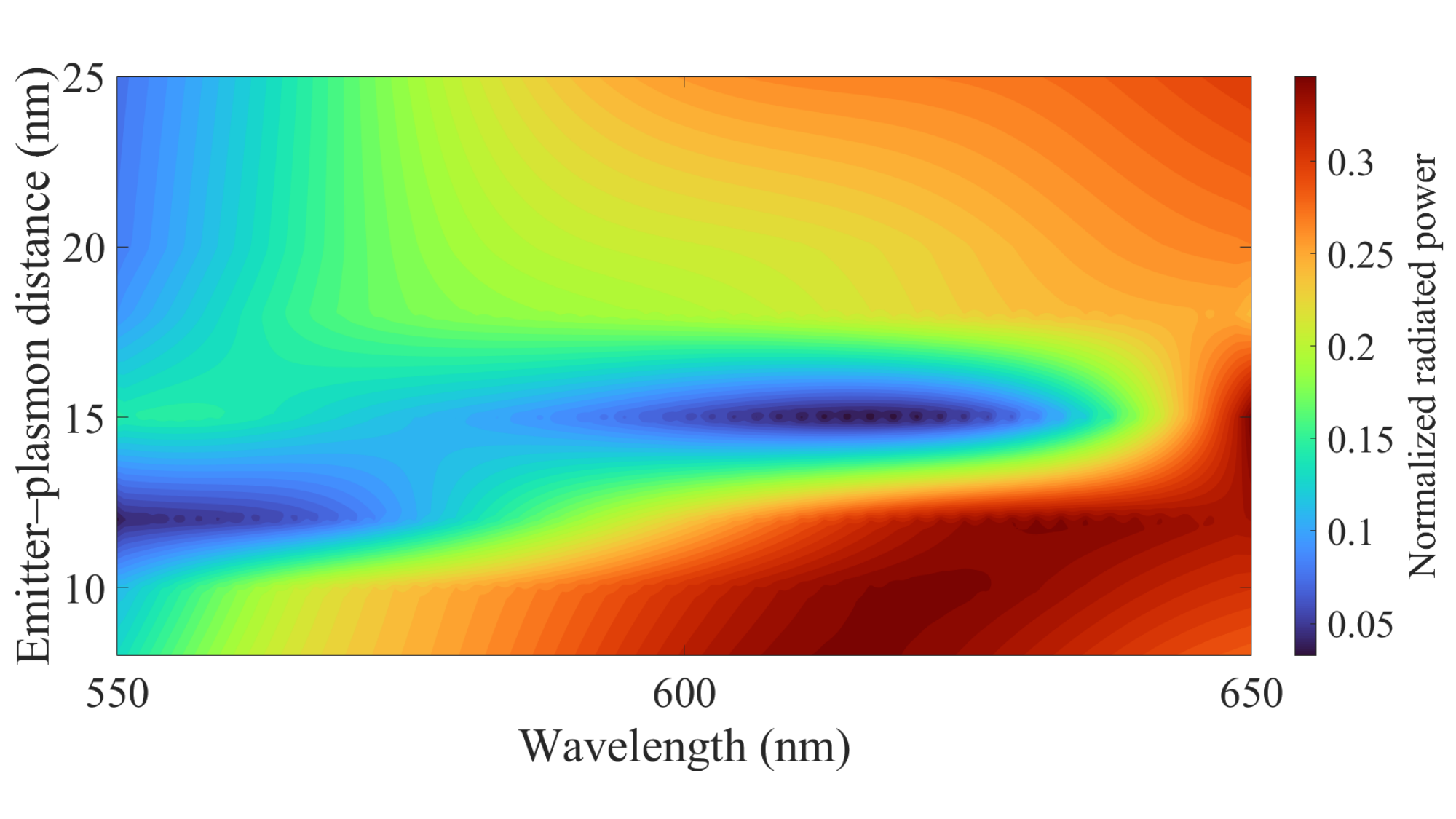}
        (b)
    \end{minipage}
  \caption{Distance and wavelength dependent radiated power for the 
  \ce{Cs3Cu2Br5}-based LED. (a) 3D surface plot showing a broader maximum 
  compared to \ce{Cs3Cu2Cl5}. (b) 2D contour map revealing a dark band at 
  14 to 16~nm and secondary enhancement at around 25~nm.}
  \label{fig:Br_radiated}
\end{figure}

Radiated power was mapped against wavelength and emitter-plasmon distance to see how near-field coupling turns into usable light output. This also clarifies why a 5~nm \ce{SiO2} shell works best. Fig.~\ref{fig:Cl_radiated} and Fig.~\ref{fig:Br_radiated} show the results for \ce{Cs3Cu2Cl5} and \ce{Cs3Cu2Br5}, respectively. Both figures include a 3D surface plot and a 2D contour map.

\textbf{\ce{Cs3Cu2Cl5}.} Figure~\ref{fig:Cl_radiated} shows the radiated power 
distribution for the \ce{Cs3Cu2Cl5}-based LED. A clear maximum appears at 
around 530 to 540~nm wavelength and 15~nm emitter-plasmon separation.

The sharp peak in Fig.~\ref{fig:Cl_radiated}(a) indicates that efficient 
near-field to far-field conversion occurs only within a narrow distance window. 
This behaviour arises from the competition between two effects. At short 
distances below 10~nm, the emitter couples strongly to the plasmonic near-field. 
However, this strong coupling also opens non-radiative decay channels through 
energy transfer to the metal. The energy is then dissipated as heat in the Ag 
core rather than being radiated to the far field. At longer distances above 
20~nm, the near-field intensity decays exponentially and coupling becomes too 
weak to provide significant enhancement.

The optimal distance of around 15~nm represents a balance between these 
competing effects. At this separation, the emitter experiences strong field 
enhancement while remaining outside the dominant quenching zone. The contour 
plot in Fig.~\ref{fig:Cl_radiated}(b) shows a bright horizontal band at this 
distance across the 520 to 560~nm wavelength range. The narrow spectral width 
of this band reflects the resonant nature of plasmonic enhancement. Maximum 
radiation occurs when the emission wavelength matches both the LSPR and the 
optimal coupling distance simultaneously.

The 5~nm \ce{SiO2} shell thickness used in this work places the perovskite 
emitter layer at approximately 10 to 20~nm from the Ag core surface, depending 
on dipole position within the active layer. This range falls within the 
optimal coupling window identified in Fig.~\ref{fig:Cl_radiated}(b), validating 
the shell design choice.

\textbf{\ce{Cs3Cu2Br5}.} Figure~\ref{fig:Br_radiated} shows the radiated power 
distribution for \ce{Cs3Cu2Br5}. The behaviour is notably different from 
\ce{Cs3Cu2Cl5}. The optimal coupling region is shifted to shorter distances 
of 8 to 12~nm and longer wavelengths of 620 to 650~nm.

The shift to shorter optimal distance can be explained by the higher refractive 
index of \ce{Cs3Cu2Br5}. The plasmonic near-field decays faster in higher-index 
media because the field penetration depth scales inversely with refractive 
index. To maintain strong coupling, the emitter must be placed closer to the 
metal surface. However, this also increases the risk of non-radiative quenching.

A dark band shows up at 14 to 16~nm in the contour plot of Fig.~\ref{fig:Br_radiated}(b). In this range, the emitter sits too far from the metal for strong plasmonic coupling, but not far enough to radiate efficiently on its own. The high refractive index still traps much of the light through total internal reflection.

A secondary enhancement region appears around 25~nm. At this distance, plasmonic coupling weakens and the emitter behaves more like a free dipole in a layered stack. Less interaction with the metal reduces non-radiative losses, allowing some light to escape through standard outcoupling rather than plasmon-mediated scattering. However, this secondary peak is broader and weaker than the primary one, indicating that plasmonic enhancement remains the more efficient extraction path.

This more complex distance dependence is why \ce{Cs3Cu2Br5} ends up with lower LEE (26\%) than \ce{Cs3Cu2Cl5} (30\%), even though their spectral overlap is similar. The tighter optimal coupling window combined with stronger optical confinement makes efficient extraction harder to maintain across the full active layer.

\textbf{Design Implications.} This distance-dependent analysis provides 
important guidance for device fabrication. For \ce{Cs3Cu2Cl5}, the optimal 
emitter-plasmon separation of around 15~nm allows some flexibility in shell 
thickness and emitter positioning. For \ce{Cs3Cu2Br5}, the tighter optimal 
window of 8 to 12~nm requires more precise control over nanostructure placement. 
These constraints should be considered when translating the simulation results 
to experimental device fabrication.

\subsection{Experimental Validation and Fabrication Outlook}

The calculated emission wavelengths agree well with experimental reports. 
\ce{Cs3Cu2Cl5} emits at 525--526~nm in photoluminescence 
measurements,\cite{Li2021FrontMater,Lian2020AdvSci} consistent with our 
simulations. PLQYs approaching 100\% have been reported for \ce{Cs3Cu2Cl5} 
and 97\% for \ce{Cs3Cu2I5} with optimized 
synthesis,\cite{Chen2025AdvMater,Han2022NanoEnergy} confirming that these 
materials are intrinsically efficient emitters.

\begin{table}[htbp]
\centering
\small
  \caption{Comparison of plasmonic enhancement in perovskite light-emitting systems.}
  \label{tbl:comparison}
  \begin{tabular}{llll}
    \toprule
    \textbf{Plasmonic structure} & \textbf{Emitter} & \textbf{Enhancement} & \textbf{Ref.} \\
    \midrule
    Ag nanocubes & \ce{CsPbBr3} QDs & 3.5$\times$ PL, 4.5$\times$ rate & \cite{Li2019JPCC} \\
    Ag nanowire network & \ce{CsPbBr3} QDs & 6$\times$ PL & \cite{Li2022MicroNano} \\
    Au NPs in PEDOT:PSS & \ce{CsPbBr3} & 3.8$\times$ PL & \cite{Xu2022MicroNano} \\
    Ag/\ce{SiO2} nanorod & \ce{Cs3Cu2Cl5} & 4.4$\times$ Purcell & This work \\
    Ag/\ce{SiO2} nanorod & \ce{Cs3Cu2Br5} & 2.8$\times$ Purcell & This work \\
    Ag/\ce{SiO2} nanosphere & \ce{Cs3Cu2I5} & 2.4$\times$ Purcell & This work \\
    \bottomrule
  \end{tabular}
\end{table}

Functional \ce{Cs3Cu2X5}-based LEDs have been demonstrated. Jun et al. 
reported \ce{Cs3Cu2I5} blue LEDs with maximum luminance of 10~cd/m$^2$ 
using thermal evaporation.\cite{Jun2018AdvMater} Liu et al. achieved 
70~cd/m$^2$ and 0.1\% EQE.\cite{Liu2020ACSApplMater} More recent work 
on mixed-phase copper halide LEDs reached 0.8\% EQE.\cite{Wang2023JCSU} 
The \ce{Cu+} oxidation state provides excellent ambient stability, with 
films maintaining optical properties for months without 
encapsulation.\cite{Lian2020AdvSci}

The predicted Purcell enhancement values are comparable to other 
plasmon-enhanced perovskite systems (Table~\ref{tbl:comparison}). Ag 
nanocube arrays coupled with perovskite quantum dots achieved 3.5-fold 
fluorescence intensity and 4.5-fold emission rate 
enhancement.\cite{Li2019JPCC} Ag nanowire networks demonstrated up to 
6-fold enhancement.\cite{Li2022MicroNano} Au nanoparticles in PEDOT:PSS 
showed 3.8-fold PL improvement in blue perovskite LEDs.\cite{Xu2022MicroNano} 
Our predicted LEE also exceeds the 10--20\% typical of standard perovskite 
LEDs.\cite{Zhao2023}

Ag/\ce{SiO2} core-shell nanoparticles have been successfully integrated 
into perovskite solar cells. He et al. showed that Ag@\ce{SiO2} with 5~nm 
shell thickness improved PCE by 16\%.\cite{He2020Nanomaterials} Wang et al. 
demonstrated Ag@\ce{SiO2} nanowires achieving 18\% PCE.\cite{Wang2017AdvElecMater} 
The \ce{SiO2} shell prevents charge transfer to the metal while preserving 
near-field coupling.

Several fabrication challenges should be noted. Precise nanoparticle 
positioning within the device stack remains difficult. Solution-processed 
deposition may yield non-uniform distribution, reducing the average 
enhancement below simulated values. Nevertheless, both \ce{Cs3Cu2X5} films 
and Ag/\ce{SiO2} nanoparticles can be synthesized using established methods, 
suggesting experimental realization is feasible.

\section{Conclusions}

An integrated DFT-FDTD framework has been developed to investigate 
plasmon-enhanced light extraction in lead-free \ce{Cs3Cu2X5} (X = Cl, Br, I) 
LEDs. First-principles calculations provided composition-specific 
optical constants that were directly implemented in device-level FDTD 
simulations for nanostructure optimization.

Among the three compositions studied, \ce{Cs3Cu2Cl5} exhibits the best 
overall performance with 4.4$\times$ Purcell enhancement, 30\% light 
extraction efficiency, and near-Lambertian far-field emission. This 
superior response originates from the lower refractive index of the 
material, which reduces dielectric screening and enables stronger 
near-field coupling with the plasmonic nanorod. \ce{Cs3Cu2Br5} achieves 
the highest spectral overlap (0.955) but yields only moderate extraction 
efficiency (26\%) due to increased optical confinement. \ce{Cs3Cu2I5} 
exhibits the weakest response with extraction efficiency limited to 10\%. 
This result demonstrates that Purcell enhancement alone does not ensure 
efficient light extraction in high-index host materials.

The distance analysis shows that optimal coupling conditions depend on composition. \ce{Cs3Cu2Cl5} has a wider tolerance window with best performance near 15~nm, while \ce{Cs3Cu2Br5} needs tighter control at 8--12~nm. These differences matter for experimental device fabrication.

Some limitations should be noted. The simulations assume high internal quantum efficiency and do not include non-radiative recombination or defect losses. The reported values therefore represent optical upper bounds. Real device performance will depend on material quality and defect passivation. Ohmic losses in the metal nanostructures also create a trade-off between radiative enhancement and absorption, which requires careful spacer design.

The results show that optical design for plasmon-enhanced lead-free LEDs needs to account for composition. The combination of chemical stability, 
high intrinsic quantum yield, and strong plasmonic coupling identifies 
\ce{Cs3Cu2Cl5} as a promising candidate for next-generation light-emitting 
applications. Experimental validation through fabrication of plasmon-integrated 
\ce{Cs3Cu2X5} devices represents the logical next step.

\section*{Author contributions}
\textbf{Shoumik Debnath}: Conceptualization, Methodology, Software, Validation, Formal analysis, Investigation, Data curation, Writing -- original draft, Writing -- review \& editing, Visualization. \textbf{Sudipta Saha}: Methodology, Software, Validation, Formal analysis, Writing -- review \& editing. \textbf{Khondokar Zahin}: Methodology, Validation, Data curation, Writing -- review \& editing. \textbf{Ying Yin Tsui}: Writing -- review \& editing, Supervision, Project administration. \textbf{Md. Zahurul Islam}: Conceptualization, Resources, Writing -- review \& editing, Supervision, Project administration.

\section*{Conflicts of interest}
There are no conflicts to declare.

\section*{Data availability}
Representative FDTD simulation files supporting this study are available on GitHub at \url{https://github.com/debnath-shoumik/PeLED_FDTD_Cs3Cu2X5}. All remaining data are available from the corresponding author upon reasonable request.

\section*{Acknowledgement}
The authors acknowledge the financial support provided by the Basic Research Grant, Bangladesh University of Engineering and Technology (BUET), under Office Order No.: Shongstha/R-60/Re-2413, dated 10 October 2023 (Professor Dr.\ Md.\ Zahurul Islam). The authors also acknowledge the logistical support provided by the Department of Electrical and Electronic Engineering (EEE), BUET, throughout the duration of this work.

\bibliography{Cu_X_manuscript}
\bibliographystyle{unsrt}

\end{document}